\documentclass[usenatbib]{mn2e}
\usepackage{rotating}
\usepackage{epsfig}
\usepackage{times}
\def\gtrsim{~\rlap{$>$}{\lower 1.0ex\hbox{$\sim$}}}
\def\ltsim{~\rlap{$<$}{\lower 1.0ex\hbox{$\sim$}}}

\title[MIPS photometry for Herschel-SPIRE Local Galaxies Programs]
  {MIPS 24-160~$\mu$m photometry for the Herschel-SPIRE Local Galaxies 
  Guaranteed Time Programs}

\author[G. J. Bendo et al.]
    {G. J. Bendo$^{1,2}$, F. Galliano$^3$, S. C. Madden$^3$ \\
    $^1$  UK ALMA Regional Centre Node, Jodrell Bank Centre for Astrophysics, 
          School of Physics and Astronomy, University of Manchester, 
          Oxford Road,\\ Manchester M13 9PL, United Kingdom\\
    $^2$  Astrophysics Group, Imperial College, Blackett Laboratory,
          Prince Consort Road, London SW7 2AZ, United Kingdom\\
    $^3$  Laboratoire AIM, CEA, Universit\'e Paris Diderot, 
          IRFU/Service d'Astrophysique, Bat. 709, 91191 Gif-sur-Yvette, 
          France\\
}
\date{}
\pagerange{\pageref{firstpage}--\pageref{lastpage}}
\pubyear{}

\begin{document}
\label{firstpage}
\maketitle

\begin{abstract}
We provide an overview of ancillary 24, 70, and 160~$\mu$m data from
the Multiband Imaging Photometer for {\it Spitzer} (MIPS) that are
intended to complement the 70-500~$\mu$m {\it Herschel} Space
Observatory photometry data for nearby galaxies obtained by the {\it
Herschel}-SPIRE Local Galaxies Guaranteed Time Programs and the {\it
Herschel} Virgo Cluster Survey. The MIPS data can be used to extend
the photometry to wave bands that are not observed in these {\it
Herschel} surveys and to check the photometry in cases where
{\it Herschel} performs observations at the same wavelengths.
Additionally, we measured globally-integrated 24-160~$\mu$m flux
densities for the galaxies in the sample that can be used for the
construction of spectral energy distributions.  Using MIPS photometry
published by other references, we have confirmed that we are obtaining
accurate photometry for these galaxies.
\end{abstract}

\begin{keywords}
    infrared: galaxies, galaxies: photometry, catalogues
\end{keywords}

\section{Introduction}

The {\it Herschel}-SPIRE Local Galaxies Guaranteed Time Programs (SAG2)
comprise several {\it Herschel} Space Observatory \citep{prpetal10}
programs that used primarily the Photodetector Array Camera and
Spectrometer \citep[PACS;][]{pwgetal10} and Spectral and Photometric
Imaging Receiver \citep[SPIRE;][]{gaaetal10} to perform far-infrared
and submillimetre observations of galaxies in the nearby universe.
Three of the programs include photometric surveys of galaxies.  The
Very Nearby Galaxies Survey (VNGS; PI: C. D. Wilson) has performed
70-500~$\mu$m photometric and spectroscopic observations of 13
archetypal nearby galaxies that includes Arp~220, M51, and M81.  The
Dwarf Galaxy Survey (DGS; PI: S. C. Madden) is a 70-500~$\mu$m
photometric and spectroscopic survey of 48 dwarf galaxies selected to
span a range of metallicities (with 12+log(O/H) values ranging from
7.2 to 8.5).  The Herschel Reference Survey \citep[HRS; ][]{becetal10}
is a 250-500~$\mu$m photometric survey of a volume-limited sample of
323 nearby galaxies designed to include both field and Virgo Cluster
galaxies.  The HRS also significantly overlaps with the {\it Herschel} Virgo
Cluster Survey \citep[HeViCS ][]{detal10a}, a 100-500~$\mu$m survey
that will image 60 square degrees of the Virgo Cluster, and both
collaborations will be sharing their data.

The far-infrared and submillimetre photometric data from these surveys
can be used to construct spectral energy distributions (SEDs) of the
dust emission and to map the distribution of cold dust within
these galaxies.  However, the surveys benefit greatly from the
inclusion of 24, 70, and 160~$\mu$m data from the Multiband Imaging Photometer
for {\it Spitzer} \citep[MIPS; ][]{retal04}, the far-infrared
photometric imager on board the {\it Spitzer} Space Telescope
\citep{wetal04}.  The 24~$\mu$m MIPS data are particularly important
either when attempting to model the complete dust emission from
individual galaxies, as it provides constraints on the hot dust
emission, or when attempting to measure accurate star formation rates,
as 24~$\mu$m emission has been shown to be correlated with other star
formation tracers \citep{cetal05, cetal07, petal07, ketal07, ketal09,
zetal08}.  The 70~$\mu$m MIPS data are less critical for the VNGS and
DGS galaxies, which have been mapped with PACS at 70~$\mu$m, but the
data are more important for the HRS galaxies, most of which will not
be mapped with PACS at 70~$\mu$m.  None the less, the MIPS 70~$\mu$m
data can be used to check the PACS photometry, and the data may be
useful as a substitute for PACS photometry in situations where the
MIPS data are able to detect emission at higher signal-to-noise levels
but where the higher resolution of PACS is not needed.  For galaxies
without 70~$\mu$m PACS observations, the MIPS data will provide an
important additional data point that is useful for constraining the
part of the far-infrared SED that represents the transition between
the $\sim20$~K dust emission from the diffuse interstellar medium and
the hot dust emission from large grains in star forming regions and
very small grains.  The 160~$\mu$m MIPS data are less important, as
160~$\mu$m PACS observations with equivalent sensitivities and smaller
PSFs have been performed on the VNGS and DGS samples as well as the
fraction of the HRS sample that falls within the HeViCS fields.  For
these galaxies, the MIPS 160~$\mu$m data can primarily be used to
check PACS 160~$\mu$m photometry. An additional follow-up program
(Completing the PACS coverage of the Herschel Reference Survey, P.I.:
L. Cortese) has been submitted to perform PACS 160~$\mu$m observations
on the HRS galaxies outside the HeViCS field.  However, those
observations have not yet been performed at the time of this writing,
so the MIPS 160~$\mu$m data can serve as a substitute for the missing
PACS data.

The pipeline processing from the MIPS archive is not optimized for
observations of individual galaxies.  The final 24~$\mu$m images may
include gradients from zodiacal light emission, incomplete
flatfielding, and foreground asteroids, while the 70 and 160~$\mu$m
images may include short-term variations in the background signal
(``drift'').  Moreover, many galaxies are often observed multiple
times in multiple Astronomical Observation Requests (AORs), and
optimal images can often be produced by combining the data from these
multiple AORs, which is something that the MIPS pipeline is not
designed to do.  Hence, to get the best MIPS images for analysis, it
is necessary to reprocess the archival data.

Work on reprocessing the archival MIPS data for the SAG2 and HeViCS
programs has been ongoing since before the launch of {\it Herschel}.
Either these reprocessed MIPS data or earlier versions of the data
have already been used in multiple papers from the SAG2 collaboration
\citep{cbbetal10, eetal10, gmgetal10, gbcetal10, oetal10, pcsetal10,
ssbetal10, asbetal11, betal12, sgeetal11, fetal12} and the HeViCS
collaboration \citep[]{dbzetal10, svbetal10, dbcetal12}, and the data
have also been used in other publications outside of these
collaboration \citep{ybl09, wetal09, wimgb09, getal09, cbidk10,
bwwetal10, dbbetal11}.  The data processing has been described with
some details in some of these papers but not in others.  Global
photometry measurements (printed numerical values, not just data
points shown in figures) have only been published for 11 galaxies, and
some of the measurements are based either on older versions of the
data processing or on images created before all of the MIPS data for
the targets were available.

The goal of this paper is to describe the MIPS data processing for
SAG2 in detail and to present photometry for all of the SAG2 galaxies
as well as the 500~$\mu$m flux-limited sample of HeViCS galaxies
published by \citet{dbcetal12}.  While the MIPS data is
incomplete for the DGS, HRS, and HeViCS samples and hence cannot be
used to create statistically complete datasets, the data are still
useful for constructing SEDs for individual galaxies and subsets of
galaxies in the SAG2 and HeViCS samples.  The paper is divided into
two primary sections.  Section~\ref{s_data} describes the data
processing in detail.  Section~\ref{s_photometry} describes the
globally-integrated photometry for these galaxies, which can be used
as a reference for other papers, and also discusses how the photometry
compares to the MIPS photometry from other surveys.

\section{Data processing}
\label{s_data}

\subsection{Overview of MIPS}

This section gives a brief overview of the MIPS instrument and the
type of data produced by the instrument.  Additional information on
the instrument and the arrays can be found in the MIPS Instrument
Handbook \citep{mips11}\footnote[18]{The MIPS Instrument Handbooks is
available at
http://irsa.ipac.caltech.edu/data/SPITZER/docs/mips/ \\ mipsinstrumenthandbook/MIPS\_Instrument\_Handbook.pdf
.}.

MIPS has four basic observing modes, but most observations were
performed in one of the two imaging modes.  The photometry map mode
produced maps of multiple dithered frames that were usually
$\sim5$~arcmin in size.  The observing mode could also be used to
produce raster maps or could be used in cluster mode to produce maps
of multiple objects that are close to each other.  Although intended
to be used for observing sources smaller than 5~arcmin, the mode was
sometimes used to image larger objects.  Because the 24, 70, and
160~$\mu$m arrays are offset from each other in the imaging plane,
observations in each wave band need to be performed in separate
pointings.  The scan map observing mode was designed to be used
primarily for observing objects larger than 5~arcmin.  The telescope
scans in a zig-zag pattern where each of the arrays in the instrument
pass over the target region in each scan leg.  In typical
observations, the telescope scans a region that is 1 degree in length,
although longer scan maps were also produced with the instrument.  In
both observing modes, a series of individual data frames are taken in
a cycle with the telescope pointing at different offsets from the
target.  These cycles include stimflash observations, which are frames
in which the arrays are illuminated with an internal calibration
source.  Between 6 and 32 frames may be taken during a photometry map
observation.  In scan map observations, the number of frames per cycle
may vary, but the data are always bracketed by stimflash frames.  In
typical 1 degree long scan map legs taken with the medium scan rate,
each scan leg contains 4 cycles of data, and each cycle contains 25
frames.

The other two observing modes were a 65-97~$\mu$m low resolution
spectroscopy mode using the 70~$\mu$m array and a total power mode
that could be used to measure the total emission from the sky.
However, since our interest is in working with photometric images of
individual galaxies, we did not use the data from either of these
observing modes.  

Details on the three arrays are given in Table~\ref{t_array}.  The
70~$\mu$m array is actually a $32 \times 32$ array, but half of the
array was effectively unusable, so the array effectively functions as
a $32 \times 16$ array.  Details on the effective viewing area are
given in the table.  Also, the 70~$\mu$m array can be used in wide
field-of-view and super-resolution modes for producing photometry
maps, but virtually no super-resolution data was taken for our target
galaxies, so we only list data for the wide field-of-view mode.

\begin{table*}
\centering
\begin{minipage}{131mm}
\caption{Data on the three MIPS arrays$^a$}
\label{t_array}
\begin{tabular}{@{}lcccccc@{}}
\hline
Wave Band &    Pixel Size &   
               \multicolumn{2}{c}{Array Size} &    
               PSF FWHM$^b$ &     Flux Conversion Factors &
               Calibration\\
($\mu$m) &     (arcsec) &
               (pixels) &         (arcmin) &
               (arcsec) &         (MJy/sr) [MIPS unit]$^{-1}$ &
               Uncertainty\\
\hline
24 &           $2.5 \times 2.6$ &
               $128 \times 128$ & $5.4 \times 5.4$ &
               6$^c$ &            $4.54\times10^{-2}$$^c$ &
               4\%$^c$ \\
70 &           $9.9 \times 10.1$ &
               $32 \times 16$ &   $5.2 \times 2.6$ &
               18$^d$ &           $702^d$ &
               10\%$^d$ \\
160 &          $16 \times 18$ &
               $2 \times 20$ &    $2.1 \times 5.3$ &
               38$^e$ &           $41.7^e$ &
               12\%$^e$ \\
\hline
\end{tabular}
$^a$ Except where noted, these data come from the MIPS Instrument Handbook
    \citep{mips11}.\\
$^b$ This is the full-width and half-maximum (FWHM) of the point spread 
    function (PSF).\\
$^c$ Data are from \citet{eetal07}.\\
$^d$ Data are from \citet{getal07}.\\
$^e$ Data are from \citet{setal07}.\\
\end{minipage}
\end{table*}

\subsection{Overview of data}

\subsubsection{Archival data}

{\it Spitzer} observations of multiple galaxies within the SAG2
samples were performed in other survey programs before SAG2 began
working on the MIPS analysis and data reduction.  The only {\it
Spitzer} observing program devoted to SAG2 photometry that was awarded
observing time was a program that included MIPS 24~$\mu$m observations
for 10 of the DGS galaxies, which is described in the next subsection.
All other MIPS data originate from an assortment of programs.  Some
galaxies were observed as specific targets in surveys of nearby
galaxies.  Others were observed in surveys of wide fields, such as the
wide field surveys of the Virgo Cluster.  Still others were
serendipitously observed in observations with other targets, such as
scan map observations of zodiacal light.  Both photometry maps and
scan map data are available for these galaxies.  Consequently, the
observed areas vary significantly among the galaxies. The coverage
(the number of data frames covering each pixel in the final mosaics)
and on-source integration times also vary among the galaxies.

Given the inhomogeneity of the data as well as the incomplete coverage
of the galaxies in the sample, we opted to use all data available for
every galaxy to produce the best images for each galaxy.  This means
that the data set will not be uniform and that the noise levels in the
data will vary among the galaxies in the sample, but the resulting
images will be the best on hand for analysis.  While we generally
attempted to use all available, we made some judgments on selecting
data for final images.  When both scan map and photometry map data
were available for individual galaxies, we used only the scan map data
to create final images if the optical discs of the galaxies were
larger than the areas covered in the photometry maps or if the
background area in the photometry map was too small to allow us to
apply data processing steps that rely on measurements from the
background in on-target frames.  We also did not use observations that
covered less than half of the optical discs of individual objects.
When multiple objects were covered in regions covered in multiple
overlapping or adjacent AORs, we made larger mosaics using all of the
data whenever technically feasible.  Also, for photometry map data, we
often used the serendipitous data taken when individual arrays were in
off-target positions if those fields covered galaxies in our samples,
and when multiple fields were observed using the cluster option in the
photometry map data \citep[see the MIPS Instrument handbook by
the][]{mips11}, we combined the data from all pointings that covered
SAG2 or HeViCS galaxies.

\subsubsection{SAG2 observations of dwarf galaxies}

Ten of the dwarf galaxies were observed by DGS with MIPS in cycle 5 as
part of the program Dust Evolution in Low-Metallicity Environments:
Bridging the Gap Between Local Universe and Primordial Galaxies (PI:
F. Galliano; ID: 50550).  Since these were objects smaller than 5
arcmin in diameter and since SAG2 intended to rely upon {\it Herschel}
for 70 and 160~$\mu$m photometry, these galaxies were mapped only at
24~$\mu$m using the photometry map mode.  One AOR was performed per
object.  Each observation uses a dither pattern to cover a $\sim6$
arcmin square region around the targets, and the integration times
were set to 3 s per frame, giving a total time of 328 s per AOR.

\subsection{Data processing for individual data frames}

The raw data from the {\it Spitzer} archive were reprocessed using the
MIPS Data Analysis Tools \citep{getal05} along with additional
processing steps, some of which are performed by software from the
MIPS Instrument Team and some of which were developed independently.
The scan map data processing is a variant of the data processing
pipeline used in the fourth data delivery of MIPS data from the {\it
Spitzer} Infrared Nearby Galaxies Survey \citep[SINGS;][]{ketal03},
although changes have been made to the background subtraction, and an
asteroid removal step has been added to the 24~$\mu$m data processing.
Although other data processing software for MIPS is available from the
{\it Spitzer} Science Center, we have continued to use the MIPS DAT
because of our familiarity with the software and because we have
developed an extensive range of tools to work with the intermediate
and final data products produced by the MIPS DAT.

Separate sections are used to describe the processing steps applied to
the 24~$\mu$m data frames and the steps applied to the 70 and
160~$\mu$m data frames, as the data from the 24~$\mu$m silicon-based
detectors differs somewhat from the data from the 70 and 160~$\mu$m
germanium-gallium detectors.  The tools for processing photometry map
data frames differ slightly from the tools for the scan map data
frames.  However, the differences are small enough that it is possible
to describe the data processing for both observing modes in the same
sections.  The mosaicking and post-processing steps applied to all
data are very similar, and so these steps are described in the last
subsection.

\subsubsection{MIPS 24~$\mu$m data frame processing}

The raw 24~$\mu$m data consist of slopes to the ramps measured by the
detectors (the counts accumulated in each pixel during non-destructive
readouts).  The following data processing steps were applied to MIPS
24~$\mu$m data frames:

\newcounter{step24}
\begin{list}{$\arabic{step24}$.}
{\usecounter{step24} \setlength{\leftmargin}{1em} \setlength{\labelwidth}{1em}}
\item The MIPS DAT program mips\_sloper was applied to the frames.
  This applies a droop correction, which removes an excess signal in
  each detector that is proportional to the signal in the entire
  array, a dark current subtraction, and an electronic nonlinearity
  correction.
\item The MIPS DAT program mips\_caler was applied to the data frames.
  This corrects the detector responsivity using a mirror-position
  dependent flatfield that removes spots from the images caused by
  material on the scan mirror.  This data processing step also
  included a correction for variations in the readout offsets between
  different columns in the data frames.
\item To remove latent images from bright sources, 
  pixels with signals above 2500 MIPS units in individual frames were 
  masked out in the following three frames.  In a few cases, this threshold
  was lowered to remove additional latent image effects.
\item When some 24~$\mu$m data frames were made, the array was hit by
  strong cosmic rays that also caused severe ``jailbar'' effects or
  background offsets in the data.  When we have identified data frames
  with these problems or other severe artefacts, we masked out those
  data frames manually at this stage in the data processing.
\item A mirror-position independent flatfield was created from
  on-target frames falling outside ``exclusion'' regions that included
  the optical disc of target galaxies and bright foreground or
  background sources.  These flatfields correct for responsivity
  variations in the array that are specific to each observation.  This
  flatfield was then applied to the data frames.  In the case of some
  photometry map data, not enough background area was available for
  properly making flatfields.  In these cases, data from the
  off-target pointings were used to build the mirror-position
  independent flatfields that were then applied to the data.
\item Gradients in the background signal, primarily from zodiacal
  light, were then subtracted from the data frames.  This step differs
  between the photometry and scan map modes. For photometry map data,
  the background signal outside the exclusion regions in each frame
  was fit with a plane, and then this plane was subtracted from the
  data (although this step was skipped if not enough area was
  available in the data frames to measure the background).  In the
  scan map data, two different approaches were used.  Before applying
  either of these methods, we typically discarded the first five
  frames of data from each scan leg because the background signal was
  often ramping up to a stable background level; these frames usually
  did not cover any targets.  In the standard approach, the background
  was subtracted in two steps.  First, the median signal for data
  outside the exclusion regions in each data frame were fit with a
  second-order polynomial that was a function of time, and then this
  function was subtracted from the data.  Second, we measured the mean
  residual background signal as a function of the frame position
  within a stimflash cycle and subtracted these background variations
  from the data.  The alternate background subtraction approach relies
  upon using data from multiple scan legs; it was generally applied
  when the standard approach did not properly subtract the background.
  It was also sometimes used in place of the standard approach on data
  that did not scan 1 degree with the medium scan rate (6.5 arcsec
  s$^{-1}$), as the code was simply more flexible to use.  For all
  forward scan leg data or all reverse scan leg data, we measured the
  median background level as a position of location within the scan
  leg.  This gives the background signal as a function of position in
  a scan leg and scan direction that is then applied to each scan leg.
  Note that these steps will also remove large scale structure outside
  of the exclusion regions from the data but do not significantly
  affect signal from compact and unresolved sources.
\item In cases where we had data from multiple AORs that overlapped
  similar regions, we compared the data from pairs of AORs to perform
  asteroids removal in a three step process that involved.  In the
  first step, we used the mips\_enhancer in the MIPS DAT to make
  preliminary mosaics of the data from each AOR.  In the second step,
  we subtracted the data from each AOR to produce difference maps in
  which asteroids and other transient sources will appear as either
  bright or dark sources but where stationary objects will appear as
  noise.  To identify locations that contained signal from asteroids,
  we looked for data where signal in either of the AORs was above a
  set S/N threshold, where the signal in the difference maps was above
  a set S/N threshold, and where the coverage was above a set
  threshold; these thresholds needed to be manually adjusted for each
  comparison.  When performing this step, we visually confirmed that
  the software was identifying transient sources and not stationary
  sources or background noise.  In the final step, we went through the
  data frames from each AOR and masked out data within 5 pixels ($\sim
  12.5$ arcsec) of pixels identified as containing signal from
  asteroids.  In cases with bright asteroids, we may identify multiple
  pixels containing signal from asteroids, and so we often masked out
  regions signficantly larger than 11 pixels.
\end{list}

\subsubsection{MIPS 70-160~$\mu$m data frame processing}

The raw 70 and 160~$\mu$m data consist of the counts accumulated in
each pixel during non-destructive readouts, which are referred to as
ramps.  We applied the following processing steps to the 70 and
160~$\mu$m data frames:

\newcounter{step70160}
\begin{list}{$\arabic{step70160}$.}
{\usecounter{step70160} \setlength{\leftmargin}{1em} 
  \setlength{\labelwidth}{1em}}
\item The MIPS DAT program mips\_sloper was applied to the individual
  data frames to convert the ramps into slopes.  This step also removes
  cosmic rays and readout jumps, and it includes a nonlinearity corrections.
\item The MIPS DAT program mips\_caler was applied to adjust the
  detector responsivity relative to the stim flashes observed during
  the observations and to apply illumination corrections.  This step
  also includes electronic nonlinearity and dark current corrections.
\item Short term drift in the signal was removed from the data on a
  pixel-by-pixel basis.  The background signal was measured in data
  falling outside the optical disc of the galaxy and other sources
  that we identified in exclusion regions similar to those described
  in the 24~$\mu$m data processing.  In the 70~$\mu$m photometry map
  data, the background was measured as a function of time and then
  subtracted from the data.  The 160~$\mu$m photometry map
  observations often did not include enough background data to perform
  this step properly, and the background variations in the 160~$\mu$m
  data was not problematic.  However, when the 160~$\mu$m photometry
  map data were to be combined with scan map data, we did measure
  median background signals in the areas outside the exclusion regions
  on a frame-by-frame basis and subtract these backgrounds from the
  data.  In the case of the scan map data, the median background
  signal was measured for each pixel during each stim flash cycle, a
  spline procedure was used to describe the background signal as a
  function of time during the entire AOR, and then this background was
  subtracted from the data.  This procedure also removes gradients and
  large-scale structure from regions outside the exclusion regions but
  will generally not affect compact and unresolved sources.
\item In scan map data, residual variations in the background signal
  as a function of time since the last stim flash were measured in data
  outside the exclusion regions and then subtracted from the data.
\item Any problematic data that we have identified, such as individual
  160~$\mu$m detector pixels with very poor drift correction over a
  subset of the data frames or cosmic ray hits on 160~$\mu$m detectors
  that were not filtered out in the previous data processing steps,
  were masked out manually.
\end{list}

\subsection{Mosaicking data and post-processing}

Final images for the galaxies were created using all suitable AORs
using the mips\_enhancer in a two step process.  In the first step,
the mips\_enhancer is used to identify pixels from individual frames
that are statistical outliers compared to co-spatial pixels from other
frames.  These pixels are then masked out in enhanced versions of the
data frames.  In the second step, the mips\_enhancer is used to create
the final maps.  In these images, north is up, east is left, and the
pixel scales are set to 1.5, 4.5, and 9.0 arcsec pixel$^{-1}$.  The
pixel scales are based on a convention originally adopted by SINGS, as
it allows for fine sampling of PSF substructure and as the pixel
scales are integer multiples of each other, which allows for easier
comparisons among the images.  

The CRPIX keywords in the final FITS images correspond to the centres
of the optical discs of the individual target galaxies as given by the
NASA/IPAC Extragalactic Database.  In cases where two or more galaxies
fell in contiguous areas, we sometimes produced separate final mosaics
for each galaxy in which the final maps were constructed using
different CRPIX values. We also attempted to do this for a large
amount of contiguous data for the Virgo Cluster covering a
$\sim5^\circ$ region centered on a point near NGC~4486 and an
overlapping $\sim2.5^\circ$ region approximately centered on
RA=12:28:10 Dec=+80:31:35.  While we succeeded at doing this with the
70 and 160~$\mu$m data, mips\_enhancer failed to execute properly when
we attempted this with the 24~$\mu$m data, probably because of the
relatively large angular area compared to the pixel size.  We
therefore produced final 24~$\mu$m mosaics of each galaxy in this
region based on subsets of the contiguous data.  In doing this, we
ensured that, when producing a 24~$\mu$m image of an individual
galaxy, we mosaicked all AORs that covered each galaxy that was being
mapped.  NGC 4380 is an exception, as it lies near the ends of a
$\sim5^\circ$ scan to the north and a $\sim2.5^\circ$ scan to the
south.  We therefore measured the 24~$\mu$m flux density for this
galaxy in the map produced for NGC~4390, which is nearby and which
falls in almost all of the scan maps centered on or to the north of
NGC~4380.  We also had problems with producing 24~$\mu$m maps of NGC
4522 with the CRPIX values set to the central coordinates of the
galaxy, so we measured the flux density in the map centered on
NGC~4519.  In the cases of NGC 3226/NGC 3227 and NGC 4567/NGC4568,
where the galaxies appear close enough that their optical discs
overlap, we only made one map with the central position set to the
centre of the galaxy that is brighter at optical wavelengths.

We performed a few post-processing steps to the final mosaics.  We
applied the flux calibration factors given in Table~\ref{t_array} to
produce maps in units of MJy sr$^{-1}$.  Next, we applied a
non-linearity correction to 70~$\mu$m pixels that exceeded 66 MJy
sr$^{-1}$.  This correction, given by \citet{dggetal07} as
\begin{equation}
f_{70\mu m}(\mbox{true})=0.581(f_{70\mu m}(\mbox{measured}))^{1.13}
\end{equation}
is based on data from \citet{getal07}.  When applying this correction,
we adjusted the calculations to include the median background signal
measured in the individual data frames before the drift removal steps.
We then measured and subtracted residual background surface
brightnesses outside the optical discs of the galaxies in regions that
did not contain any nearby, resolved galaxies (regardless of whether
they were detected in the MIPS bands) or point-like sources.  In the
case of the 24~$\mu$m data, we used multiple small circular regions
around the centres of targets.  For the 70 and 160~$\mu$m images, we
used whenever possible two or more regions that were as large as or
larger than the optical discs of the target galaxies and that
straddled the optical disc of the galaxy.  In some of the smaller
photometry maps, however, we could not often do this, so we made our
best effort to measure the background levels within whatever
background regions were observed.  In cases where multiple galaxies
fall within the final mosaics, we only performed this background
subtraction for the central galaxy, although when performing
photometry on the other galaxies in these fields, we measured the
backgrounds in the same way around the individual targets.

The final images have a few features and artefacts that need to be
taken into consideration when using the data.  First of all, the large
scale structure outside of the target galaxies in the images has been
mostly removed.  Although the images, particularly the 160~$\mu$m
images, may contain some cirrus structure, most of the large scale
features in the cirrus have been removed.  Second, all scan map data
may contain some residual striping.  Additionally, the 70~$\mu$m
images for bright sources are frequently affected by latent image
effects that manifest themselves as positive or negative streaks
aligned with the scan direction.  Finally, many objects falling within
the Virgo Cluster as well as a few objects in other fields were
observed in fields covered only with MIPS scan map data taken using
the fast scan rate.  The resulting 160~$\mu$m data contain large gaps
in the coverage, and the data appear more noisy than most other
160~$\mu$m data because of the poor sampling.

\section{Photometry}
\label{s_photometry}

\subsection{Description of measurements}
\label{s_photometry_meas}

For most galaxies, we performed aperture photometry within elliptical
apertures with major and minor axes that were the greater of either
1.5 times the axis sizes of the D$_{25}$ isophotes given by
\citep{detal91} or 3~arcmin.  The same apertures were used in all
three bands for consistency.  The lower limit of 3 arcmin on the
measurement aperture dimensions ensures that we can measure the total
flux densities of 160~$\mu$m sources without needing to apply aperture
corrections.  We performed tests with measuring some unresolved
sources in the DGS with different aperture sizes and found that the
fraction of the total flux not included within a 3~arcmin aperture for
these sources is below the 12\% calibration uncertainty of the
160~$\mu$m band.  In galaxies much larger than 3~arcmin, we found that
apertures that were 1.5 times the D$_{25}$ isophote contained all of
the measurable signal from the target galaxies.  The measured flux
densities in apertures larger than this did not change significantly,
but the measured flux densities decreased if we used smaller
apertures.

For the elliptical galaxies NGC~3640, NGC~4125, NGC~4365, NGC~4374,
NGC~4406, NGC~4472, NGC~4486, NGC~4552, NGC~4649, NGC~4660, and
NGC~5128, however, we used measurement apertures that were the same
size as the D$_{25}$ isophotes.  Additionally, for the nearby dwarf
elliptical galaxy NGC~205, we used a measurement aperture that was 0.5
times the size of the D$_{25}$ isophote. These were all cases where
the 70 and 160~$\mu$m emission across most of the optical disc is
within $5\sigma$ of the background noise, and in many cases, the
emission from the galaxies is not detected.  Using smaller apertures
in these specific cases allows us to avoid including background
sources and artefacts from the data processing, thus allowing us to
place better constraints on the flux densities.  We also treated
NGC~4636 as a special case in which, at 160~$\mu$m, we only measured
the flux density for the central source because of issues with
possible background sources falling within the optical disc of the
galaxy (although the background sources are not as problematic at
24~$\mu$m, and so the 24~$\mu$m measurement is still for the entire
optical disc).  Additional details on NGC~4636 are given in
Section~\ref{s_photom_notes}.

A few galaxies in the various samples are so close to each other or so
close to other galaxies at equivalent distances that attempting to
separate the infrared emission from the different sources would be
very difficult.  Objects where this is the case are Mrk 1089 (within
NGC~1741), NGC 3395/3396, NGC 4038/4039, NGC~4567/4568, NGC 5194/5195,
and UM 311 (within NGC~450).  In these cases, we used measurement
apertures that were large enough to encompass the emission from the
target galaxy and all other nearby sources.  Details on the other
apertures are given in Table~\ref{t_specialaperture}.

\begin{table}
\caption{Special measurement apertures}
\label{t_specialaperture}
\begin{center}
\begin{tabular}{@{}lcccc@{}}
\hline
Galaxy &         R.A. &        Dec. &       Axis sizes &       Position \\
&               (J2000) &      (J2000)&     (arcmin) &         Angle$^a$\\
\hline 
Mrk 1089 &      05:01:37.8 &   -04:15:28 &  $3.0\times3.0$ &   $0^\circ$ \\
NGC 891 &       02:22:33.4 &   +42:20:57 &  $20.3\times10.0$ & $22^\circ$ \\
NGC 3395/3396 & 10:49:50.1 &   +32:58:58 &  $6.0\times6.0$ &   $0^\circ$ \\
NGC 4038/4039 & 12:01:53.0 &   -18:52:10 &  $10.4\times10.4$ & $0^\circ$ \\
NGC 4567/4568 & 12:36:34.3 &   +11:14:20 &  $8.5\times8.5$ &   $0^\circ$ \\
NGC 5194/5195 & 13:29:52.7 &   +47:11:43 &  $19.6\times19.6$ & $0^\circ$ \\
NGC 6822 &      19:44:56.6 &   -14:47:21 &  $30.0\times30.0$ & $0^\circ$ \\
UM 311 &        01:15:30.4 &   -00:51:39 &  $4.7\times3.5$ &   $72^\circ$ \\
\hline
\end{tabular}
\end{center}
$^a$ Position angle is defined as degrees from north through east.
\end{table}

Many of the galaxies in the DGS do not have optical discs defined by
\citet{detal91}, and some do not have optical discs defined anywhere in the
literature.  These are generally galaxies smaller than the minimum
3~arcmin diameter aperture that we normally use, so we used
measurement apertures of that size in many cases.  However, for
sources fainter than 100 mJy in the 24~$\mu$m data, we found that
background noise could become an issue when measuring 24~$\mu$m flux
densities over such large apertures; although the galaxy would clearly
be detected at a level much higher than $5 \sigma$ in the centre of
the aperture, the integral of the aperture would make the detection
appear weaker.  Hence, for 24~$\mu$m DGS sources that were fainter
than 10 mJy and did not appear extended in the 24~$\mu$m data, we
used apertures with 1~arcmin diameters and divided the data by 0.93,
which is an aperture correction that we derived empirically from
bright point-like sources in the DGS.

NGC~891 and NGC~6822 were treated as special cases for selecting the
measurement apertures.  Details are given in the photometry notes
below, and the parameters describing the measurement apertures are
given in Table~\ref{t_specialaperture}.

Before performing the photometry on individual galaxies, we identified
and masked out emission that appeared to be unrelated to the target
galaxies.  We visually identified and masked out artefacts from the
data processing in the final mosaics, such as bright or dark pixels
near the edges of mapped field and streaking in the 70~$\mu$m images
related to latent image effects.  We also statistically checked for
pixels that were $5\sigma$ below the background, which are almost
certainly associated with artefacts except when this becomes
statistically probable in apertures containing large numbers of
pixels.  In cases where we determined that the $<-5\sigma$ pixels were
data processing artefacts or excessively noisy pixels, we masked them
out.  When other galaxies appeared close to individual galaxies in
which we were measauring flux densities but when the optical discs did
not overlap significantly, we masked out the adjacent galaxies. We
also masked out emission from unresolved sources, particularly
unresolved 24~$\mu$m sources, that did not appear to be associated
with the target galaxies and that appeared signficantly brighter than
the emission in the regions where we measured the background.  Most of
these sources appeared between the D$_{25}$ isophote and the
measurement aperture.  In cases where the galaxies contained very
compact 24~$\mu$m emission (as is the case for many elliptical and S0
galaxies), we also masked out unresolved sources within but near the
D$_{25}$ isophote.  A few unresolved sources within the D$_{25}$
isophote appeared as bright, unresolved sources in Digitized Sky
Survey or 2MASS data, indicating that they were foreground stars, and
we masked them out as well.  In many 24~$\mu$m images, the measured
flux densities changed by less than 4\% (the calibration uncertainty)
when the unresolved sources were removed.

As stated above, in cases where the MIPS 160~$\mu$m data for
individual galaxies consists of only scan map data taken at the fast
scan rate, our final 160~$\mu$m maps include gaps in the coverage.  To
make 160~$\mu$m measurements, we have interpolated the signal across
these gaps using nearest neighbor sampling techniques.  We also
applied this interpolation technique to 160~$\mu$m data for the
regions in the optical discs (but not in the whole measurement
aperture, which may fall outside the scan region) of IC 1048, NGC
4192, NGC 4535, and NGC 5692.  In many other cases, the observed
regions did not completely cover the optical discs of the target
galaxies.  We normally measured the flux densities for the regions
covered in the observed regions.  Cases where the observed regions did
not cover $\gtrsim90$\% of the optical discs are noted in the
photometry tables.  Although we believe that these data are reliable
(especially since the observations appear to cover most of the
emission that is seen in the other bands), people using these data
should still be aware of the limitations of these data.

As a quality check on the photometry, we examined the 24/70, 24/160,
and 70/160~$\mu$m flux density ratios to identify any galaxies that
may have discrepant colours (for example, abnormally high 24/70 and
low 70/160~$\mu$m colours, which would be indicative of problems with
unmasked negative pixels in the final images).  In such discrepant
cases, we examined the images for unmasked artefacts, masked out the
artefacts when identified, and repeated the photometry.

The globally-integrated flux densities for the galaxies in the four
different samples are listed in
Tables~\ref{t_photom_vngs}-\ref{t_photom_hevicsextra}.  No colour
corrections have been applied to these data.  We include three sources
of uncertainty.  The first source is the calibration uncertainty.  The
second source is the uncertainty based on the error map.  Each pixel
in the error map is based on the standard deviation of the overlapping
pixels from the individual data frames; the uncertainties will include
both instrumental background noise and shot noise from the
astronomical sources.  To calculate the total uncertainty traced by
the data in the error map, we used the square root of the sum of the
square of the error map pixels in the measurement region.  The third
source of uncertainty is from background noise (which includes both
instrumental and astronomical sources of noise) measured in the
background regions.  The total uncertainties are calculated by adding
these three sources of uncertainty in quadrature.  Sources that are
less than $5\sigma$ detections within the measurement apertures
compared to the combination of the error map and background noise are
reported as $5\sigma$ upper limits.  Sources in which the surface
brightness within the measurement aperture is not detected at the
$5\sigma$ level for regions unaffected by foreground/background
sources or artefacts are reported as upper limits; in these cases, the
integrated flux densities within the apertures are used as upper
limits.  This second case occurs when the target aperture includes
emission from diffuse, extended emission (as described for NGC~4552
below) or large scale artefacts that are impossible to mask out for
the photometry.

\begin{table*}
\footnotesize
\centering
\begin{minipage}{175mm}
\caption{Photometry for the Very Nearby Galaxies Survey}
\label{t_photom_vngs}
\begin{tabular}{@{}lcccccccccc@{}}
\hline
Galaxy &    
               \multicolumn{4}{c}{Optical Disc} &
	       Wavelength &
	       Flux Density &
               \multicolumn{4}{c}{Flux Density Uncertainty (Jy)$^d$}\\
 &
	       R. A. &          Declination &
	       Axes &           Position &
               ($\mu$m) &
               Measurement & 
               Calibration &    Error &
               Background &     Total \\
 &
               (J2000)$^a$ &    (J2000)$^a$ &
	       (arcmin)$^b$ &   Angle$^{bc}$ &
	       &
               (Jy) & 
               &                Map &
               &                \\               
\hline 
NGC 205  &  00:40:22.0 &  +41:41:07 & $21.9\times11.0$ &   $170^\circ$ &
            24 &  0.1089 &  0.0044 &  0.0005 &  0.0008 &  0.0044 \\
& & & & &   70 &   1.302 &   0.130 &   0.019 &   0.023 &  0.134 \\
& & & & &   160 &   8.98 &    1.08 &    0.03 &    0.05 &  1.08 \\
NGC 891$^e$ &  02:22:33.4 &  +42:20:57 &  $13.5\times2.5$ &   $22^\circ$ & 
            24 &  6.4531 &  0.2581 &  0.0005 &  0.0007 &  0.2581 \\
& & & & &   70 &  97.122 &   9.712 &   0.045 &   0.018 &  9.712 \\
& & & & &   160 & 287.27 &   34.47 &    8.72 &    0.04 &  35.56 \\
NGC 1068 &  02:42:40.7 &  -00:00:48 &  $7.1\times6.0$ &    $70^\circ$ &
            24 &         &         &         &         &  \\
& & & & &   70 & 189.407 &  18.941 &  0.491 &  0.058 &  18.947 \\
& & & & &   160 & 237.39 &  28.49 &  5.53 &  0.06 &  29.02 \\
NGC 2403 &  07:36:51.4 &  +65:36:09 &  $21.9\times12.3$ &  $127^\circ$ &
            24 &  6.0161 &  0.2406 &  0.0022 &  0.0019 &  0.2407 \\
& & & & &   70 &  81.710 &   8.171 &   0.057 &  0.052 &  8.171 \\
& & & & &   160 & 221.04 &  26.53 &  0.24 &  0.11 &  26.53 \\
NGC 3031 &  09:55:33.1 &  +69:03:55 &  $26.9\times14.1$ &  $157^\circ$ &
            24 &  5.2748 &  0.2110 &  0.0017 &  0.0024 &  0.2110 \\
& & & & &   70 &  81.049 &   8.105 &   0.063 &  0.080 &  8.106 \\
& & & & &   160 & 316.30 &  37.96 &  0.97 &  0.40 &  37.97 \\
NGC 4038$^f$ & & &  &  &
            24 &  5.8226 &  0.2329 &  0.0073 &  0.0012 &  0.2330 \\
& & & & &   70 &  45.949 &   4.595 &   0.148 &  0.035 &  4.597 \\
& & & & &   160 &  80.28 &   9.63 &   3.62 &  0.06 &  10.29 \\
NGC 4125 &  12:08:06.0 &  +65:10:27 &  $5.8\times3.2$ &    $95^\circ$ & 
            24 &  0.0790 &  0.0032 &  0.0002 &  0.0003 &  0.0032\\
& & & & &   70 &   1.014 &    0.101 &   0.008 &  0.008 &  0.102 \\
& & & & &   160 &   1.37 &    0.16 &   0.01 &  0.01 &  0.17 \\
NGC 4151 &  12:10:32.5 &  +39:24:21 &  $6.3\times4.5$ &    $50^\circ$ &
            24 &  4.5925 &  0.1837 &  0.0104 &  0.0005 &  0.1840 \\
& & & & &   70 &   5.415 &    0.541 &   0.027 &  0.013 &  0.542 \\
& & & & &   160 &   9.38 &    1.13 &   0.02 &  0.02 &  1.13 \\
NGC 5128 &  13:25:27.6 &  -43:01:09 &  $25.7\times20.0$ &  $35^\circ$ &
            24 &  24.0374 & 0.9615 &  0.0135 &  0.0028 &  0.9616 \\
& & & & &   70 & 263.165 &  26.316 &  0.226 &  0.068 &  26.318 \\
& & & & &   160 & 582.51 &  69.90 &  22.50 & 0.14 &  73.43 \\
NGC 5194$^f$ & & &  &  &
            24 &  14.2309 & 0.5692 &  0.0037 &  0.0015 &  0.5693 \\
& & & & &   70 &  151.000 &  15.100 &  0.123 &  0.045 &  15.101 \\
& & & & &   160 & 458.44 &  55.01 &  7.80 &  0.11 &  55.56 \\
NGC 5236 &  13:37:00.9 &  -29:51:57 &  12.9 &               &
            24 &  40.4266 & 1.6171 &  0.0263 &  0.0017 &  1.6173 \\
& & & & &   70 & 312.808 &  31.281 &  0.290 &  0.051 &  31.282 \\
& & & & &   160 & 798.23 &  95.78 &  9.95 &  0.13 &  96.30 \\
Arp 220  &  15:34:57.1 &  +23:30:11 &   1.5 &              &
            24 &         &         &         &         &  \\
& & & & &   70 &  74.976 &   7.498 &   0.309 &   0.023 &  7.504 \\ 
& & & & &   160 &  54.88 &    6.59 &    1.38 &    0.02 &  6.73 \\
\hline
\end{tabular}
$^a$ Data are from NED.\\
$^b$ Data are from \citet{detal91} unless otherwise specified.  If
\citet{detal91} specify both the minor/major axis ratio 
and the position angle, then both axes and the position angle are
listed.  If \citet{detal91} did not specify either of these data, then
we performed photometry on circular regions, and so only the major
axis is specified.\\
$^c$ The position angle is defined as degrees from north through east.\\
$^d$ Details on the sources of these uncertainties are given in
Section~\ref{s_photometry_meas}.\\
$^e$ A special measurement aperture was used for NGC 891.  See
Table~\ref{t_specialaperture}.\\
$^f$ These objects consist of two galaxies with optical discs that
overlap.  See Table~\ref{t_specialaperture} for the dimensions of the
measurement apertures for these objects.\\
\end{minipage}
\end{table*}

\begin{table*}
\footnotesize
\centering
\begin{minipage}{171mm}
\caption{Photometry for the Dwarf Galaxies Survey}
\label{t_photom_dgs}
\begin{tabular}{@{}lcccccccccc@{}}
\hline
Galaxy &    
               \multicolumn{4}{c}{Optical Disc} &
	       Wavelength &
	       Flux Density &
               \multicolumn{4}{c}{Flux Density Uncertainty (Jy)$^d$}\\
 &
	       R. A. &          Declination &
	       Axes &           Position &
               ($\mu$m) &
               Measurement & 
               Calibration &    Error &
               Background &     Total \\
 &
               (J2000)$^a$ &    (J2000)$^a$ &
	       (arcmin)$^b$ &   Angle$^bc$ &
	       &
               (Jy) & 
               &                Map &
               &                \\               
\hline 
IC 10        &  00:20:17.3 &  +59:18:14 &  6.3 &    &
      24 &   9.8188 &   0.3928 &  0.0136 &  0.0013 &  0.3930 \\
& & & & &   70 &       &          &        &        &        \\
& & & & &   160 &           &       &       &       &       \\    
HS 0017+1055 &  00:20:21.4 &  +11:12:21 &  &                  &
            24 &   0.0237 &   0.0009 &  0.0005 &  0.0009 &  0.0014 \\
& & & & &   70 &       &          &        &        &        \\
& & & & &   160 &           &       &       &       &       \\    
Haro 11      &  00:36:52.4 &  -33:33:19 &  &                  &
            24 &   2.3046 &   0.0922 &  0.0123 &  0.0005 &  0.0930 \\
& & & & &   70 & 4.912 &    0.491 &  0.038 &  0.007 &  0.493 \\
& & & & &   160 &      2.01 &  0.24 &  0.01 &  0.02 &  0.24 \\    
HS 0052+2536 &  00:54:56.3 &  +25:53:08 &  &                  &
            24 &   0.0207 &   0.0008 &  0.0004 &  0.0008 &  0.0012 \\
& & & & &   70 &       &          &        &        &        \\
& & & & &   160 &           &       &       &       &       \\    
UM 311$^e$       & & & & & 
            24 &   0.3289 &   0.0132 &  0.0008 &  0.0009 &  0.0132 \\
& & & & &   70 & 3.075 &    0.308 &  0.008 &  0.008 &  0.308 \\
& & & & &   160 &      6.62 &  0.79 &  0.02 &  0.01 &  0.79 \\    
NGC 625      &  01:35:04.6 &  -41:26:10 &  $5.8\times1.9$ &   $92^\circ$ &
    24 &   0.8631 &   0.0345 &  0.0016 &  0.0003 &  0.0346 \\
& & & & &   70 & 6.252 &    0.625 &  0.036 &  0.012 &  0.626 \\
& & & & &   160 &      7.87 &  0.94 &  0.03 &  0.02 &  0.95 \\    
UGCA 20      &  01:43:14.7 &  +19:58:32 &  $3.1\times0.8$ &   $153^\circ$ &
    24 &   $<0.0085$ &       &         &         &         \\
& & & & &   70 &       &          &        &        &        \\
& & & & &   160 &           &       &       &       &       \\    
UM 133       &  01:44:41.2 &  +40:53:26 &  &                  &
            24 &   0.0094 &   0.0004 &  0.0002 &  0.0003 &  0.0005 \\
& & & & &   70 &       &          &        &        &        \\
& & & & &   160 &           &       &       &       &       \\    
UM 382       &  01:58:09.3 &  -00:06:38 &  &                  &
            24 &          &          &         &         &         \\
& & & & &   70 & $<0.070$ &       &        &        &        \\
& & & & &   160 &           &       &       &       &       \\    
NGC 1140     &  02:54:33.5 &  -10:01:40 &  $1.7\times0.9$ &   $10^\circ$ &
    24 &   0.3764 &   0.0151 &  0.0009 &  0.0006 &  0.0151 \\
& & & & &   70 & 3.507 &    0.351 &  0.020 &  0.008 &  0.351 \\
& & & & &   160 &      3.67 &  0.44 &  0.01 &  0.01 &  0.44 \\    
SBS 0335-052 &  03:37:44.0 &  -05:02:40 &  &                  &
            24 &   0.0768 &   0.0031 &  0.0005 &  0.0005 &  0.0032 \\
& & & & &   70 & 0.051 &    0.005 &  0.005 &  0.006 &  0.009 \\
& & & & &   160 &   $<0.07$ &       &       &       &       \\    
NGC 1569     &  04:30:49.0 &  -64:50:53 &  $3.6\times1.8$ &   $120^\circ$ &
    24 &   7.7189 &   0.3088 &  0.0091 &  0.0010 &  0.3089 \\
& & & & &   70 & 46.120 &   4.612 &  0.068 &  0.029 &  4.613 \\
& & & & &   160 &     33.49 &  4.02 &  0.11 &  0.02 &  4.02 \\    
NGC 1705     &  04:54:13.5 &  -53:21:40 &  $1.9\times1.4$ &   $50^\circ$ &
    24 &   0.0532 &   0.0021 &  0.0000 &  0.0001 &  0.0021 \\
& & & & &   70 & 1.315 &    0.132 &  0.002 &  0.004 &  0.132 \\
& & & & &   160 &      1.29 &  0.16 &  0.01 &  0.01 &  0.16 \\    
Mrk 1089$^e$     &   &   &  &  &
      24 &   0.5252 &   0.0210 &  0.0008 &  0.0003 &  0.0210 \\
& & & & &   70 & 1.123 &    0.112 &  0.004 &  0.004 &  0.112 \\
& & & & &   160 &           &       &       &       &       \\    
II Zw 40     &  05:55:42.6 &  +03:23:32 &  &                  &
            24 &   1.6545 &   0.0662 &  0.0063 &  0.0006 &  0.0665 \\
& & & & &   70 & 5.438 &    0.544 &  0.031 &  0.011 &  0.545 \\
& & & & &   160 &  &  &  &  &  \\    
Tol 0618-402 &  06:20:02.5 &  -40:18:09 &  &                  &
            24 &   $<0.0015$ &       &         &         &         \\
& & & & &   70 & $<0.037$ &       &        &        &        \\
& & & & &   160 &   $<0.42$ &       &       &       &       \\    
NGC 2366     &  07:28:54.6 &  +69:12:57 &  $8.1\times3.3$ &   $25^\circ$ &
    24 &   0.6919 &   0.0277 &  0.0013 &  0.0007 &  0.0277 \\
& & & & &   70 & 5.230 &    0.523 &  0.021 &  0.019 &  0.524 \\
& & & & &   160 &      5.50 &  0.66 &  0.21 &  0.03 &  0.69 \\    
HS 0822+3542 &  08:25:55.5 &  +35:32:32 &  &                  &
            24 &   0.0032 &   0.0001 &  0.0001 &  0.0002 &  0.0003 \\
& & & & &   70 & 0.043 &    0.004 &  0.004 &  0.006 &  0.008 \\
& & & & &   160 &   $<0.04$ &       &       &       &       \\    
He 2-10      &  08:36:15.1 &  -26:24:34 &  &                  &
            24 &   5.7368 &   0.2295 &  0.0262 &  0.0007 &  0.2310 \\
& & & & &   70 & 17.969 &   1.797 &  0.102 &  0.009 &  1.800 \\
& & & & &   160 &     13.41 &  1.61 &  0.05 &  0.01 &  1.61 \\    
UGC 04483    &  08:37:03.0 &  +69:46:31 &  &                  &
            24 &   0.0101 &   0.0004 &  0.0001 &  0.0003 &  0.0005 \\
& & & & &   70 & 0.142 &    0.014 &  0.003 &  0.006 &  0.016 \\
& & & & &   160 &      0.27 &  0.03 &  0.01 &  0.00 &  0.03 \\    
\hline
\end{tabular}
\end{minipage}
\end{table*}
\addtocounter{table}{-1}

\begin{table*}
\centering
\begin{minipage}{171mm}
\caption{Photometry for the Dwarf Galaxies Survey (continued)}
\begin{tabular}{@{}lcccccccccc@{}}
\hline
Galaxy &    
               \multicolumn{4}{c}{Optical Disc} &
	       Wavelength &
	       Flux Density &
               \multicolumn{4}{c}{Flux Density Uncertainty (Jy)$^d$}\\
 &
	       R. A. &          Declination &
	       Axes &           Position &
               ($\mu$m) &
               Measurement & 
               Calibration &    Error &
               Background &     Total \\
 &
               (J2000)$^a$ &    (J2000)$^a$ &
	       (arcmin)$^b$ &   Angle$^bc$ &
	       &
               (Jy) & 
               &                Map &
               &                \\               

\hline
I Zw 18      &  09:34:02.0 &  +55:14:28 &  &                  &
            24 &   0.0061 &   0.0002 &  0.0001 &  0.0002 &  0.0003 \\
& & & & &   70 & 0.042 &    0.004 &  0.002 &  0.004 &  0.006 \\
& & & & &   160 &   $<0.12$ &       &       &       &       \\    
Haro 2       &  10:32:31.9 &  +54:24:03 &  &   &
            24 &   0.8621 &   0.0345 &  0.0015 &  0.0001 &  0.0345 \\
& & & & &   70 & 3.988 &    0.399 &  0.019 &  0.005 &  0.399 \\
& & & & &   160 &      3.09 &  0.37 &  0.01 &  0.01 &  0.37 \\    
Haro 3       &  10:45:22.4 &  +55:57:37 &  &                  &
            24 &   0.8514 &   0.0341 &  0.0027 &  0.0004 &  0.0342 \\
& & & & &   70 & 4.898 &    0.490 &  0.018 &  0.007 &  0.490 \\
& & & & &   160 &      3.93 &  0.47 &  0.01 &  0.01 &  0.47 \\    
Mrk 153      &  10:49:05.0 &  +52:20:08 &  &                  &
            24 &   0.0358 &   0.0014 &  0.0003 &  0.0005 &  0.0015 \\
& & & & &   70 & 0.260 &    0.026 &  0.004 &  0.007 &  0.027 \\
& & & & &   160 &           &       &       &       &       \\    
VII Zw 403   &  11:27:59.8 &  +78:59:39 &  &                  &
            24 &   0.0329 &   0.0013 &  0.0002 &  0.0005 &  0.0014 \\
& & & & &   70 & 0.425 &    0.043 &  0.005 &  0.007 &  0.043 \\
& & & & &   160 &      0.31 &  0.04 &  0.00 &  0.01 &  0.04 \\    
Mrk 1450     &  11:38:35.6 &  +57:52:27 &  &                  &
            24 &   0.0570 &   0.0023 &  0.0003 &  0.0004 &  0.0023 \\
& & & & &   70 & 0.264 &    0.026 &  0.004 &  0.005 &  0.027 \\
& & & & &   160 &      0.15 &  0.02 &  0.00 &  0.01 &  0.02 \\    
UM 448       &  11:42:12.4 &  +00:20:03 &  &                  &
            24 &   0.6425 &   0.0257 &  0.0018 &  0.0007 &  0.0258 \\
& & & & &   70 & 3.703 &    0.370 &  0.021 &  0.015 &  0.371 \\
& & & & &   160 &      2.67 &  0.32 &  0.01 &  0.01 &  0.32 \\    
UM 461       &  11:51:33.3 &  -02:22:22 &  &                  &
            24 &   0.0344 &   0.0014 &  0.0002 &  0.0029 &  0.0032 \\
& & & & &   70 & 0.090 &    0.009 &  0.003 &  0.011 &  0.014 \\
& & & & &   160 &      0.10 &  0.01 &  0.00 &  0.01 &  0.01 \\    
SBS 1159+545 &  12:02:02.3 &  +54:15:50 &  &                  &
            24 &   0.0062 &   0.0002 &  0.0001 &  0.0002 &  0.0004 \\
& & & & &   70 &       &          &        &        &        \\
& & & & &   160 &           &       &       &       &       \\    
SBS 1211+540 &  12:14:02.4 &  +53:45:17 &  &                  &
            24 &   0.0033 &   0.0001 &  0.0001 &  0.0002 &  0.0003 \\
& & & & &   70 &       &          &        &        &        \\
& & & & &   160 &           &       &       &       &       \\    
NGC 4214     &  12:15:39.1 &  +36:19:37 &  8.5 & &
      24 &   2.1044 &   0.0842 &  0.0015 &  0.0012 &  0.0842 \\
& & & & &   70 & 24.049 &   2.405 &  0.043 &  0.032 &  2.406 \\
& & & & &   160 &     38.18 &  4.58 &  0.34 &  0.05 &  4.59 \\    
Tol 1214-277 &  12:17:17.0 &  -28:02:33 &  &                  &
            24 &   0.0068 &   0.0003 &  0.0001 &  0.0002 &  0.0003 \\
& & & & &   70 & 0.073 &    0.007 &  0.004 &  0.005 &  0.010 \\
& & & & &   160 &      &   &   &   &   \\    
HS 1222+3741 &  12:24:36.7 &  +37:24:37 &  &                  &
            24 &          &          &         &         &         \\
& & & & &   70 & 0.062 &    0.006 &  0.004 &  0.007 &  0.010 \\
& & & & &   160 &           &       &       &       &       \\    
Mrk 209      &  12:26:16.0 &  +48:29:37 &  &                  &
            24 &   0.0587 &   0.0023 &  0.0003 &  0.0005 &  0.0024 \\
& & & & &   70 & 0.466 &    0.047 &  0.004 &  0.004 &  0.047 \\
& & & & &   160 &      0.18 &  0.02 &  0.00 &  0.01 &  0.02 \\    
NGC 4449     &  12:28:11.8 &  +44:05:40 &  $6.2\times4.4$ &   $45^\circ$ &
    24 &   3.2863 &   0.1315 &  0.0010 &  0.0008 &  0.1315 \\
& & & & &   70 & 43.802 &   4.380 &  0.053 &  0.019 &  4.381 \\
& & & & &   160 &     78.09 &  9.37 &  0.70 &  0.03 &  9.40 \\    
SBS 1249+493 &  12:51:52.4 &  +49:03:28 &  &                  &
            24 &   0.0043 &   0.0002 &  0.0001 &  0.0002 &  0.0003 \\
& & & & &   70 &       &          &        &        &        \\
& & & & &   160 &           &       &       &       &       \\    
NGC 4861     &  12:59:02.3 &  +34:51:34 &  $4.0\times1.5$ &   $15^\circ$ &
    24 &   0.3657 &   0.0146 &  0.0012 &  0.0008 &  0.0147 \\
& & & & &   70 & 1.971 &    0.197 &  0.012 &  0.010 &  0.198 \\
& & & & &   160 &      2.00 &  0.24 &  0.01 &  0.02 &  0.24 \\    
HS 1304+3529 &  13:06:24.1 &  +35:13:43 &  &                  &
            24 &   0.0122 &   0.0005 &  0.0004 &  0.0007 &  0.0009 \\
& & & & &   70 &       &          &        &        &        \\
& & & & &   160 &           &       &       &       &       \\    
Pox 186      &  13:25:48.6 &  -11:36:38 &  &                  &
            24 &   0.0108 &   0.0004 &  0.0005 &  0.0009 &  0.0011 \\
& & & & &   70 &       &          &        &        &        \\
& & & & &   160 &           &       &       &       &       \\    
NGC 5253     &  13:39:55.9 &  -31:38:24 &  $5.0\times1.9$ &   $45^\circ$ &
    24 &          &          &         &         &         \\
& & & & &   70 & 23.626 &   2.363 &  0.074 &  0.015 &  2.364 \\
& & & & &   160 &     17.35 &  2.08 &  0.05 &  0.03 &  2.08 \\    
\hline
\end{tabular}
\end{minipage}
\end{table*}
\addtocounter{table}{-1}

\begin{table*}
\centering
\begin{minipage}{171mm}
\caption{Photometry for the Dwarf Galaxies Survey (continued)}
\begin{tabular}{@{}lcccccccccc@{}}
\hline
Galaxy &    
               \multicolumn{4}{c}{Optical Disc} &
	       Wavelength &
	       Flux Density &
               \multicolumn{4}{c}{Flux Density Uncertainty (Jy)$^d$}\\
 &
	       R. A. &          Declination &
	       Axes &           Position &
               ($\mu$m) &
               Measurement & 
               Calibration &    Error &
               Background &     Total \\
 &
               (J2000)$^a$ &    (J2000)$^a$ &
	       (arcmin)$^b$ &   Angle$^bc$ &
	       &
               (Jy) & 
               &                Map &
               &                \\               

\hline
SBS 1415+437 &  14:17:01.3 &  +43:30:05 &  &                  &
            24 &   0.0187 &   0.0007 &  0.0003 &  0.0005 &  0.0009 \\
& & & & &   70 & 0.177 &    0.018 &  0.004 &  0.006 &  0.019 \\
& & & & &   160 &   $<0.06$ &       &       &       &       \\    
HS 1424+3836 &  14:26:28.1 &  +38:22:59 &  &                  &
            24 &          &          &         &         &         \\
& & & & &   70 & $<0.024$ &       &        &        &        \\
& & & & &   160 &           &       &       &       &       \\    
HS 1442+4250 &  14:44:12.8 &  +42:37:44 &  &                  &
            24 &   0.0066 &   0.0003 &  0.0001 &  0.0001 &  0.0003 \\
& & & & &   70 & 0.079 &    0.008 &  0.004 &  0.006 &  0.010 \\
& & & & &   160 &   $<0.10$ &       &       &       &       \\    
SBS 1533+574 &  15:34:13.8 &  +57:17:06 &  &                  &
            24 &          &          &         &         &         \\
& & & & &   70 & 0.270 &    0.027 &  0.004 &  0.005 &  0.028 \\
& & & & &   160 &           &       &       &       &       \\    
NGC 6822$^f$  &  19:44:56.6 &  -14:47:21 &  15.5 &  &
      24 &   4.5230 &   0.1809 &  0.0027 &  0.0032 &  0.1810 \\
& & & & &   70 & 52.413 &   5.241 &  0.082 &  0.096 &  5.243 \\
& & & & &   160 &    109.44 & 13.13 &  0.61 &  0.20 & 13.15 \\    
Mrk 930      &  23:31:58.2 &  +28:56:50 &  &                  &
            24 &   0.1985 &   0.0079 &  0.0005 &  0.0006 &  0.0080 \\
& & & & &   70 & 1.159 &    0.116 &  0.007 &  0.006 &  0.116 \\
& & & & &   160 &      0.96 &  0.12 &  0.01 &  0.02 &  0.12 \\    
HS 2352+2733 &  23:54:56.7 &  +27:49:59 &  &                  &
            24 &   0.0026 &   0.0001 &  0.0001 &  0.0003 &  0.0003 \\
& & & & &   70 &       &          &        &        &        \\
& & & & &   160 &           &       &       &       &       \\    
\hline
\end{tabular}
$^a$ Data are from NED.\\
$^b$ Data are from \citet{detal91} unless otherwise specified.  If
\citet{detal91} specify both the minor/major axis ratio 
and the position angle, then both axes and the position angle are
listed.  If \citet{detal91} did not specify either of these data, then
we performed photometry on circular regions, and so only the major
axis is specified.  If no optical dimensions are specified, then
we performed photometry on a 3~arcmin diameter circular region centered on
the source\\
$^c$ The position angle is defined as degrees from north through east.\\
$^d$ Details on the sources of these uncertainties are given in
Section~\ref{s_photometry_meas}.\\
$^e$ Special measurement apertures were used for these targets because of
the presence of nearby associated sources.  See 
Table~\ref{t_specialaperture}.\\
$^f$ A special measurement aperture was used for NGC 6822.  See
Table~\ref{t_specialaperture}.\\
\end{minipage}
\end{table*}

\begin{table*}
\footnotesize
\centering
\begin{minipage}{174mm}
\caption{Photometry for the Herscher Reference Survey}
\label{t_photom_hrs}
\begin{tabular}{@{}lccccccccccc@{}}
\hline
Galaxy &       HRS &
               \multicolumn{4}{c}{Optical Disc} &
	       Wave- &
	       Flux Density &
               \multicolumn{4}{c}{Flux Density Uncertainty (Jy)$^e$}\\
 &             Number$^a$ &
	       R. A. &          Declination &
	       Axes &           Position &
               length &
               Measurement & 
               Calibration &    Error &
               Back- &     Total \\
 &             &
               (J2000)$^b$ &    (J2000)$^b$ &
	       (arcmin)$^c$ &   Angle$^{cd}$ &
	       ($\mu$m) &
               (Jy) & 
               &                Map &
               ground &         \\               
\hline
NGC 3226 & 3 & 
            10:23:27.4 &  +19:53:55 &  $3.2\times2.8$ &  $15^\circ$ &
            24 &     0.0250 &    0.0010 &    0.0003 &    0.0006 &    0.0012 \\
& & & & & & 70 &     0.459 &    0.046 &    0.009 &    0.011 &    0.048 \\
& & & & & & 160 & & & & & \\
NGC 3227 & 4 & 
            10:23:30.5 &  +19:51:54 &  $5.4\times3.6$ &  $155^\circ$ &
            24 &     1.7173 &    0.0687 &    0.0067 &    0.0010 &    0.0690 \\
& & & & & & 70 &     9.033 &    0.903 &    0.044 &    0.018 &    0.905 \\
\hline
\end{tabular}
\end{minipage}
\end{table*}
\addtocounter{table}{-1}

\begin{table*}
\footnotesize
\centering
\begin{minipage}{174mm}
\caption{Photometry for the Herscher Reference Survey (continued)}
\begin{tabular}{@{}lccccccccccc@{}}
\hline
Galaxy &       HRS &
               \multicolumn{4}{c}{Optical Disc} &
	       Wave- &
	       Flux Density &
               \multicolumn{4}{c}{Flux Density Uncertainty (Jy)$^e$}\\
 &             Number$^a$ &
	       R. A. &          Declination &
	       Axes &           Position &
               length &
               Measurement & 
               Calibration &    Error &
               Back- &     Total \\
 &             &
               (J2000)$^b$ &    (J2000)$^b$ &
	       (arcmin)$^c$ &   Angle$^{cd}$ &
	       ($\mu$m) &
               (Jy) & 
               &                Map &
               ground &         \\               
\hline
NGC 3683 & 56 &
            11:27:31.8 &  +56:52:37 &  $1.9\times0.7$ &  $128^\circ$ &
            24 &     1.1755 &    0.0470 &    0.0003 &    0.0001 &    0.0470 \\
& & & & & & 70 & & & & & \\
& & & & & & 160 & & & & & \\
NGC 3686 & 57 & 
            11:27:43.9 &  +17:13:27 &  $3.2\times2.5$ &  $15^\circ$ &
            24 &     0.5463 &    0.0219 &    0.0004 &    0.0004 &    0.0219 \\
& & & & & & 70 & & & & & \\
\hline
\end{tabular}
\end{minipage}
\end{table*}
\addtocounter{table}{-1}

\begin{table*}
\footnotesize
\centering
\begin{minipage}{174mm}
\caption{Photometry for the Herscher Reference Survey (continued)}
\begin{tabular}{@{}lccccccccccc@{}}
\hline
Galaxy &       HRS &
               \multicolumn{4}{c}{Optical Disc} &
	       Wave- &
	       Flux Density &
               \multicolumn{4}{c}{Flux Density Uncertainty (Jy)$^e$}\\
 &             Number$^a$ &
	       R. A. &          Declination &
	       Axes &           Position &
               length &
               Measurement & 
               Calibration &    Error &
               Back- &     Total \\
 &             &
               (J2000)$^b$ &    (J2000)$^b$ &
	       (arcmin)$^c$ &   Angle$^{cd}$ &
	       ($\mu$m) &
               (Jy) & 
               &                Map &
               ground &         \\               
\hline
NGC 4298 & 111 & 
            12:21:32.7 &  +14:36:22 &  $3.2\times1.8$ &  $140^\circ$ &
            24 &     0.5129 &    0.0205 &    0.0002 &    0.0003 &    0.0205 \\
& & & & & & 70 &     5.672 &    0.567 &    0.010 &    0.008 &    0.567 \\
& & & & & & 160 &    19.18 &    2.30 &    0.03 &    0.01 &    2.30 \\
NGC 4302 & 113 & 
            12:21:42.4 &  +14:35:54 &  $5.5\times1.0$ &  $178^\circ$ &
            24 &     0.4855 &    0.0194 &    0.0002 &    0.0004 &    0.0194 \\
& & & & & & 70 &     6.286 &    0.629 &    0.014 &    0.010 &    0.629 \\
\hline
\end{tabular}
\end{minipage}
\end{table*}
\addtocounter{table}{-1}

\begin{table*}
\footnotesize
\centering
\begin{minipage}{174mm}
\caption{Photometry for the Herscher Reference Survey (continued)}
\begin{tabular}{@{}lccccccccccc@{}}
\hline
Galaxy &       HRS &
               \multicolumn{4}{c}{Optical Disc} &
	       Wave- &
	       Flux Density &
               \multicolumn{4}{c}{Flux Density Uncertainty (Jy)$^e$}\\
 &             Number$^a$ &
	       R. A. &          Declination &
	       Axes &           Position &
               length &
               Measurement & 
               Calibration &    Error &
               Back- &     Total \\
 &             &
               (J2000)$^b$ &    (J2000)$^b$ &
	       (arcmin)$^c$ &   Angle$^{cd}$ &
	       ($\mu$m) &
               (Jy) & 
               &                Map &
               ground &         \\               
\hline
IC 3322A & 143 & 
            12:25:42.5 &  +07:13:00 &  $3.5\times0.4$ &  $157^\circ$ &
            24 &     0.1816 &    0.0073 &    0.0007 &    0.0009 &    0.0074 \\
& & & & & & 70 &     2.872 &    0.287 &    0.023 &    0.024 &    0.289 \\
& & & & & & 160 &     7.97$^g$ &    0.96 &    0.04 &    0.05 &    0.96 \\
NGC 4388 & 144 & 
            12:25:46.7 &  +12:39:44 &  $5.6\times1.3$ &  $92^\circ$ &
            24 &     2.5714 &    0.1029 &    0.0041 &    0.0004 &    0.1029 \\
& & & & & & 70 &    10.730 &    1.073 &    0.029 &    0.010 &    1.073 \\
\hline
\end{tabular}
\end{minipage}
\end{table*}
\addtocounter{table}{-1}

\begin{table*}
\footnotesize
\centering
\begin{minipage}{174mm}
\caption{Photometry for the Herscher Reference Survey (continued)}
\begin{tabular}{@{}lccccccccccc@{}}
\hline
Galaxy &       HRS &
               \multicolumn{4}{c}{Optical Disc} &
	       Wave- &
	       Flux Density &
               \multicolumn{4}{c}{Flux Density Uncertainty (Jy)$^e$}\\
 &             Number$^a$ &
	       R. A. &          Declination &
	       Axes &           Position &
               length &
               Measurement & 
               Calibration &    Error &
               Back- &     Total \\
 &             &
               (J2000)$^b$ &    (J2000)$^b$ &
	       (arcmin)$^c$ &   Angle$^{cd}$ &
	       ($\mu$m) &
               (Jy) & 
               &                Map &
               ground &         \\               
\hline
NGC 4445 & 167 & 
            12:28:15.9 &  +09:26:10 &  $2.6\times0.5$ &  $106^\circ$ &
            24 &     0.0366 &    0.0015 &    0.0006 &    0.0008 &    0.0018 \\
& & & & & & 70 &     0.538 &    0.054 &    0.012 &    0.017 &    0.058 \\
& & & & & & 160 &     1.62$^g$ &    0.19 &    0.02 &    0.07 &    0.21 \\
UGC 7590 & 168 & 
            12:28:18.7 &  +08:43:46 &  $1.3\times0.4$ &  $168^\circ$ &
            24 &     0.0320 &    0.0013 &    0.0005 &    0.0007 &    0.0015 \\
& & & & & & 70 &     0.528 &    0.053 &    0.015 &    0.019 &    0.058 \\
\hline
\end{tabular}
\end{minipage}
\end{table*}
\addtocounter{table}{-1}

\begin{table*}
\footnotesize
\centering
\begin{minipage}{174mm}
\caption{Photometry for the Herscher Reference Survey (continued)}
\begin{tabular}{@{}lccccccccccc@{}}
\hline
Galaxy &       HRS &
               \multicolumn{4}{c}{Optical Disc} &
	       Wave- &
	       Flux Density &
               \multicolumn{4}{c}{Flux Density Uncertainty (Jy)$^e$}\\
 &             Number$^a$ &
	       R. A. &          Declination &
	       Axes &           Position &
               length &
               Measurement & 
               Calibration &    Error &
               Back- &     Total \\
 &             &
               (J2000)$^b$ &    (J2000)$^b$ &
	       (arcmin)$^c$ &   Angle$^{cd}$ &
	       ($\mu$m) &
               (Jy) & 
               &                Map &
               ground &         \\               
\hline
IC 797 & 189 & 
            12:31:54.7 &  +15:07:26 &  $1.3\times0.9$ &  $108^\circ$ &
            24 &     0.0731 &    0.0029 &    0.0002 &    0.0003 &    0.0029 \\
& & & & & & 70 &     1.105 &    0.111 &    0.006 &    0.010 &    0.111 \\
& & & & & & 160 &    2.31 &    0.28 &    0.01 &    0.02 &    0.28 \\
NGC 4501 & 190 &
            12:31:59.2 &  +14:25:14 &  $6.9\times3.7$ &  $140^\circ$ &
            24 &     2.2216 &    0.0889 &    0.0004 &    0.0006 &    0.0889 \\
& & & & & & 70 &    28.284 &    2.828 &    0.024 &    0.016 &    2.829 \\
\hline
\end{tabular}
\end{minipage}
\end{table*}
\addtocounter{table}{-1}

\begin{table*}
\footnotesize
\centering
\begin{minipage}{174mm}
\caption{Photometry for the Herscher Reference Survey (continued)}
\begin{tabular}{@{}lccccccccccc@{}}
\hline
Galaxy &       HRS &
               \multicolumn{4}{c}{Optical Disc} &
	       Wave- &
	       Flux Density &
               \multicolumn{4}{c}{Flux Density Uncertainty (Jy)$^e$}\\
 &             Number$^a$ &
	       R. A. &          Declination &
	       Axes &           Position &
               length &
               Measurement & 
               Calibration &    Error &
               Back- &     Total \\
 &             &
               (J2000)$^b$ &    (J2000)$^b$ &
	       (arcmin)$^c$ &   Angle$^{cd}$ &
	       ($\mu$m) &
               (Jy) & 
               &                Map &
               ground &         \\               
\hline
NGC 4567 & 215 & & & & & 
            24 &     2.0960 &    0.0838 &    0.0019 &    0.0020 &    0.0839 \\
\ \ \ /4568$^h$&/216 
& & & & & 70 &    27.005 &    2.700 &    0.109 &    0.054 &    2.703 \\
& & & & & & 160 &    68.02$^g$ &    8.16 &    0.49 &    0.05 &    8.18 \\
NGC 4569 & 217 & 
            12:36:49.7 &  +13:09:46 &  $9.5\times4.4$ &  $23^\circ$ &
            24 &     1.4279 &    0.0571 &    0.0019 &    0.0009 &    0.0572 \\
& & & & & & 70 &    11.451 &    1.145 &    0.041 &    0.021 &    1.146 \\
\hline
\end{tabular}
\end{minipage}
\end{table*}
\addtocounter{table}{-1}

\begin{table*}
\footnotesize
\centering
\begin{minipage}{174mm}
\caption{Photometry for the Herscher Reference Survey (continued)}
\begin{tabular}{@{}lccccccccccc@{}}
\hline
Galaxy &       HRS &
               \multicolumn{4}{c}{Optical Disc} &
	       Wave- &
	       Flux Density &
               \multicolumn{4}{c}{Flux Density Uncertainty (Jy)$^e$}\\
 &             Number$^a$ &
	       R. A. &          Declination &
	       Axes &           Position &
               length &
               Measurement & 
               Calibration &    Error &
               Back- &     Total \\
 &             &
               (J2000)$^b$ &    (J2000)$^b$ &
	       (arcmin)$^c$ &   Angle$^{cd}$ &
	       ($\mu$m) &
               (Jy) & 
               &                Map &
               ground &         \\               
\hline
NGC 4654 & 247 & 
            12:43:56.5 &  +13:07:36 &  $4.9\times2.8$ &  $128^\circ$ &
            24 &     1.6726 &    0.0669 &    0.0005 &    0.0004 &    0.0669 \\
& & & & & & 70 &    19.503 &    1.950 &    0.021 &    0.012 &    1.950 \\
& & & & & & 160 &    48.60 &    5.83 &    0.07 &    0.02 &    5.83 \\
NGC 4660 & 248 & 
            12:44:31.9 &  +11:11:26 &  $2.2\times1.6$ &  $100^\circ$ &
            24 &     0.0173 &    0.0007 &    0.0001 &    0.0001 &    0.0007 \\
& & & & & & 70 & $<0.072$ & & & & \\
\hline
\end{tabular}
\end{minipage}
\end{table*}
\addtocounter{table}{-1}

\begin{table*}
\footnotesize
\centering
\begin{minipage}{174mm}
\caption{Photometry for the Herscher Reference Survey (continued)}
\begin{tabular}{@{}lccccccccccc@{}}
\hline
Galaxy &       HRS &
               \multicolumn{4}{c}{Optical Disc} &
	       Wave- &
	       Flux Density &
               \multicolumn{4}{c}{Flux Density Uncertainty (Jy)$^e$}\\
 &             Number$^a$ &
	       R. A. &          Declination &
	       Axes &           Position &
               length &
               Measurement & 
               Calibration &    Error &
               Back- &     Total \\
 &             &
               (J2000)$^b$ &    (J2000)$^b$ &
	       (arcmin)$^c$ &   Angle$^{cd}$ &
	       ($\mu$m) &
               (Jy) & 
               &                Map &
               ground &         \\               
\hline
NGC 5363 & 306 &
            13:56:07.2 &  +05:15:17 &  $4.1\times2.6$ &  $135^\circ$ &
            24 &     0.1421 &    0.0057 &    0.0004 &    0.0007 &    0.0057 \\
& & & & & & 70 &     2.167 &    0.217 &    0.030 &    0.018 &    0.220 \\
& & & & & & 160 &     4.54 &    0.54 &    0.12 &    0.04 &    0.56 \\
NGC 5576 & 312 & 
            14:21:03.6 &  +03:16:16 &  $3.5\times2.2$ &  $95^\circ$ &
            24 &     0.0270 &    0.0011 &    0.0003 &    0.0006 &    0.0013 \\
& & & & & & 70 & $<0.096$ & & & & \\
\hline
\end{tabular}
$^a$ The HRS number corresponds to the numbers given by \citet{becetal10}.\\
$^b$ Data are from NED.\\
$^c$ Data are from \citet{detal91} unless otherwise specified.  If
\citet{detal91} specify both the minor/major axis ratio 
and the position angle, then both axes and the position angle are
listed.  If \citet{detal91} did not specify either of these data, then
we performed photometry on circular regions, and so only the major
axis is specified.\\
$^d$ The position angle is defined as degrees from north through east.\\
$^e$ Details on the sources of these uncertainties are given in
Section~\ref{s_photometry_meas}.\\
$^f$ These measurements are from data in which significant portions of
the optical discs ($>10$\%) of the galaxies were not covered in this
specific wave band.  The measurements here are for the region that was
covered in the MIPS data.  We have applied no corrections for the
missing flux density.\\
$^g$ These 160~$\mu$m measurements are for galaxies that were covered
in scan map observations in which the final 160~$\mu$m images for
these galaxies contain NaN values within the optical disc as a
consequence of incomplete coverage.  This typically occurs when scan
maps are performed using the fast scan rate, although NaN values
within the optical discs of galaxies occasionally appear in other
data.  The 160~$\mu$m measurements for these galaxies is based upon
interpolating over these pixels; see the text for details.\\
$^h$ These objects consist of two galaxies with optical discs that
overlap.  See Table~\ref{t_specialaperture} for the dimensions of the
measurement apertures for these objects.\\
\end{minipage}
\end{table*}

\begin{table*}
\footnotesize
\centering
\begin{minipage}{171mm}
\caption{Photometry for additional Herschel Virgo Cluster Survey galaxies$^a$}
\label{t_photom_hevicsextra}
\begin{tabular}{@{}lcccccccccc@{}}
\hline
Galaxy &    
               \multicolumn{4}{c}{Optical Disc} &
	       Wavelength &
	       Flux Density &
               \multicolumn{4}{c}{Flux Density Uncertainty (Jy)$^e$}\\
 &
	       R. A. &          Declination &
	       Axes &           Position &
               ($\mu$m) &
               Measurement & 
               Calibration &    Error &
               Background &     Total \\
 &
               (J2000)$^b$ &    (J2000)$^b$ &
	       (arcmin)$^c$ &   Angle$^{cd}$ &
	       &
               (Jy) & 
               &                Map &
               &                \\               
\hline 
NGC 4165 &
            12:12:11.7 &  +13:14:47 &  $1.3\times0.9$ &  $160^\circ$ &
            24 &     0.0264 &    0.0011 &    0.0003 &    0.0004 &    0.0012 \\
& & & & &   70 & & & & & \\
& & & & &   160 & & & & & \\
NGC 4234 &
            12:17:09.1 &  +03:40:59 &  1.3 &    &
            24 &     0.1547 &    0.0062 &    0.0002 &    0.0003 &    0.0062 \\
& & & & &   70 & & & & & \\
& & & & &   160 & & & & & \\
NGC 4252 & 
            12:18:30.8 &  +05:33:34 &  $1.5\times0.4$ &  $48^\circ$ &
            24 &     0.0098 &    0.0004 &    0.0002 &    0.0002 &    0.0005 \\
& & & & &   70 &     0.183 &    0.018 &    0.005 &    0.005 &    0.020 \\  
& & & & &   160 &     0.46 &    0.06 &    0.00 &    0.02 &    0.06 \\
NGC 4266 & 
            12:19:42.3 &  +05:32:18 &  $2.0\times0.4$ &  $76^\circ$ &
            24 &     0.0329 &    0.0013 &    0.0004 &    0.0005 &    0.0015 \\
& & & & &   70 &     0.494 &    0.049 &    0.007 &    0.011 &    0.051 \\  
& & & & &   160 &     2.04 &    0.24 &    0.01 &    0.02 &    0.25 \\
NGC 4273 & 
            12:19:56.0 &  +05:20:36 &  $2.3\times1.5$ &  $10^\circ$ &
            24 &     1.0295 &    0.0412 &    0.0011 &    0.0005 &    0.0412 \\
& & & & &   70 &    12.387 &    1.239 &    0.030 &    0.013 &    1.239 \\  
& & & & &   160 &    18.50 &    2.22 &    0.03 &    0.02 &    2.22 \\
NGC 4299 &
            12:21:40.9 &  +11:30:12 &  $1.7\times1.6$ &  $26^\circ$ &
            24 &     0.2350 &    0.0094 &    0.0002 &    0.0002 &    0.0094 \\
& & & & &   70 &     3.346 &    0.335 &    0.008 &    0.006 &    0.335 \\  
& & & & &   160 &     4.32 &    0.52 &    0.02 &    0.01 &    0.52 \\
NGC 4309 &
            12:22:12.3 &  +07:08:40 &  $1.9\times1.1$ &  $85^\circ$ &
            24 &     0.0620 &    0.0025 &    0.0003 &    0.0003 &    0.0025 \\
& & & & &   70 & & & & & \\
& & & & &   160 & & & & & \\
IC 3258 & 
            12:23:44.4 &  +12:28:42 &  1.6 &    &
            24 &     0.0764 &    0.0031 &    0.0005 &    0.0007 &    0.0032 \\
& & & & &   70 &     0.776 &    0.078 &    0.014 &    0.019 &    0.081 \\  
& & & & &   160 &     0.87 &    0.10 &    0.01 &    0.02 &    0.11 \\
NGC 4411 & 
            12:26:30.1 &  +08:52:20 &  2.0 &    &
            24 &     0.0234 &    0.0009 &    0.0005 &    0.0006 &    0.0012 \\
& & & & &   70 &     0.474 &    0.047 &    0.014 &    0.017 &    0.052 \\
& & & & &   160 &     1.40 &    0.17 &    0.01 &    0.01 &    0.17 \\
UGC 7557 &
            12:27:11.0 &  +07:15:47 &  3.0 &    &
            24 &     0.0326 &    0.0013 &    0.0007 &    0.0010 &    0.0018 \\
& & & & &   70 &     0.659 &    0.066 &    0.020 &    0.029 &    0.075 \\
& & & & &   160 &    1.55$^f$ &    0.19 &    0.03 &    0.04 &    0.19 \\
NGC 4466 &
            12:29:30.5 &  +07:41:47 &  $1.3\times0.4$ &  $101^\circ$ &
            24 &     0.0243 &    0.0010 &    0.0005 &    0.0007 &    0.0013 \\
& & & & &   70 &     0.602 &    0.060 &    0.014 &    0.020 &    0.065 \\
& & & & &   160 &     1.13$^f$ &    0.14 &    0.01 &    0.02 &    0.14 \\
IC 3476 & 
            12:32:41.8 &  +14:03:02 &  $2.1\times1.8$ &  $30^\circ$ &
            24 &     0.1881 &    0.0075 &    0.0006 &    0.0007 &    0.0076 \\
& & & & &   70 &     1.961 &    0.196 &    0.016 &    0.019 &    0.198 \\
& & & & &   160 &     2.88$^f$ &    0.35 &    0.01 &    0.02 &    0.35 \\
NGC 4531 &
            12:34:15.8 &  +13:04:31 &  $3.1\times2.0$ &  $155^\circ$ &
            24 &     0.0351 &    0.0014 &    0.0006 &    0.0009 &    0.0018 \\
& & & & &   70 &     0.539 &    0.054 &    0.017 &    0.023 &    0.061 \\
& & & & &   160 &     2.76$^f$ &    0.33 &    0.02 &    0.04 &    0.33 \\
\hline
\end{tabular}
$^a$ These are galaxies that are not in the HRS but that appear in
the 500~$\mu$m-selected sample published by \citet{dbcetal12}.\\
$^b$ Data are from NED.\\
$^c$ Data are from \citet{detal91} unless otherwise specified.  If
\citet{detal91} specify both the minor/major axis ratio 
and the position angle, then both axes and the position angle are
listed.  If \citet{detal91} did not specify either of these data, then
we performed photometry on circular regions, and so only the major
axis is specified.\\
$^d$ The position angle is defined as degrees from north through east.\\
$^e$ Details on the sources of these uncertainties are given in
Section~\ref{s_photometry_meas}.\\
$^f$ These 160~$\mu$m measurements are for galaxies that were covered
in scan map observations in which the final 160~$\mu$m images for
these galaxies contain NaN values within the optical disc as a
consequence of incomplete coverage.  This typically occurs when scan
maps are performed using the fast scan rate, although NaN values
within the optical discs of galaxies occasionally appear in other
data.  The 160~$\mu$m measurements for these galaxies is based upon
interpolating over these pixels; see the text for details.
\end{minipage}
\end{table*}

\subsubsection{Notes on photometry}
\label{s_photom_notes}

Aside from typical issues described above with the data processing and
photometry, we encountered multiple problems that were unique to
individual targets.  Notes on these issues (in the order in which the
galaxies appear in the table) are listed below.

~

{\em \noindent Notes on the VNGS data}

{\em Arp 220} - The centre of the galaxy, which is unresolved in the
MIPS bands, saturated the 24~$\mu$m detector, and so no 24~$\mu$m flux
density is reported for the source.  The 160~$\mu$m error contains two
anomalously high pixels (pixels with error map values at least an
order of magnitude higher than the image map values) located off the
peak of the emission.  We ascertained that the corresponding image map
pixels did not look anomalous compatred to adjacent pixels, so the
unusually high values in the error map were probably some type of
artefact of the data reduction possibly related to a combination of
high surface brightness issues and coverage issues.  We therefore
excluded these pixels when calculating the error map uncertainty.

{\em NGC 891} - This is an edge-on spiral galaxy in which the central
plane is very bright, and so features that look similar to Airy rings
(except that they are linear rather than ring-shaped) appear above and
below the plane of the galaxy in the 160~$\mu$m image.  The
measurement aperture we used for all three bands has a major axis
corresponding to 1.5 times the D$_{25}$ isophote but a much broader
minor axis that encompasses the vertically-extended emission.  Note
that this is the only edge-on galaxy where we have encountered this
problem.

{\em NGC 1068} - This is another galaxy that is unresolved in the MIPS
bands and that saturatesd the 24~$\mu$m detector.  It is not practical
to perform 24~$\mu$m photometry measurements on this galaxy.  The
160~$\mu$m error contains a few anomalously high pixels (pixels with
error map values at least an order of magnitude higher than the image
map values).  This seemed similar to the phenomenon described for the
anomalous 160~$\mu$m error map pixels for Arp 220.  We excluded these
pixels when calculating the error map uncertainty.

{\em NGC 3031} - The 160~$\mu$m image includes residual cirrus
emission between the D$_{25}$ isophote and the measurement aperture
that was masked out when calculating the 160~$\mu$m flux density.  See
\citet{sgmggh10} and \citet{detal10b} for details on the features.

{\em NGC 3034} - The galaxy saturates the MIPS detectors in all three
bands and causes unusually severe artefacts to appear in the data, and
so we report no photometric measurements for this galaxy.

{\em NGC 4038/4039} - The 70~$\mu$m image is strongly affected by streaking
from latent image effects.

{\em NGC 5128} - The centre of the galaxy produced latent image
effects that appear as a broad streak in the final image.  The artefact
was masked out when photometry was performed.

{\em NGC 5236} - The central 8~arcsec of the galaxy saturated the
24~$\mu$m and 160~$\mu$m data, but this region appears to contribute a
relatively small fraction of the total emission from NGC~5236.  We
think the 24~$\mu$m measuements should still be reliable to within the
calibration uncertainty of 4\%.  As for the 160~$\mu$m image,
we interpolated across the single central saturated pixel to estimate
the flux density for the pixel; the correction is much smaller than
the calibration uncertainty.

~

{\em \noindent Notes on the DGS data}

{\em HS 0052+2536} - The 24~$\mu$m image shows an unresolved 24~$\mu$m
source at the central position of HS~0052+2536 and an unresolved
24~$\mu$m source with a similar surface brightness at the central
position of HS~0052+2537, which is located $\sim15$~arcsec to the
north. We masked out HS~0052+2537 when performing photometry.

{\em IC 10} - This galaxy was observed with MIPS only in the
photometry map mode.  However, the photometry map mode is intended for
objects smaller than 5~arcmin, while the optical disc of IC~10 and the
infrared emission from it are much more extended than this.  While
$\gtrsim90$ \% of the optical disc was covered at 24~$\mu$m, only part
of the galaxy was observed at 70 and 160~$\mu$m, and a significant
fraction of the infrared emission may have fallen outside the observed
regions.  Given this, we will not report 70 and 160~$\mu$m
measurements for this galaxy.

{\em Mrk~153} - In the 160~$\mu$m image, the galaxy becomes blended
with another galaxy to the east.  We therefore do not report 160~$\mu$m flux
densities for this galaxy.

{\em NGC~5253} - This is another case where the galaxy is unresolved
in the MIPS bands and where the galaxy saturated the 24~$\mu$m
detector, which is why we report no 24~$\mu$m flux density for this
galaxy.

{\em NGC~6822} - The galaxy has an extension to the south \citep{cetal06}
that is not included within the optical disk given by \citet{detal91},
so for photometry, we used a 30~arcmin diameter circle centered on the
optical position of the galaxy given by NED.  This galaxy also lies in
a field with cirrus structure on the same size as the galaxy.  The
version of the 70 and 160~$\mu$m data processing that we applied has
removed the gradient in the cirrus emission present in this part of
the sky, which causes the final map to appear significantly different
from the SINGS version of the map for this specific galaxy.

{\em SBS 1249+493} - The 24~$\mu$m image includes a bright central
source and a fainter source $\sim12$~arcsec to the south.  It is
unclear as to whether this source is associated with the galaxy;
we masked it out before performing flux density measurements.

{\em Tol 0618-402} - The brightest feature in the 160~$\mu$m
photometry map image is a streak-like feature running from northwest
to southeast near the location of the galaxy.  It is unclear from this
image alone if this is an artefact of the data processing or a real
large-scale feature, although based on what we have seen in similar
160~$\mu$m photometry map data, the latter may be more likely.  No
feature in the image appears to correspond to the source itself, and
so we reported the integrated 160~$\mu$m flux density within the
3~arcmin diameter aperture on the source as the upper limit on the
emission, using regions flanking this region as the best background
measurements available.

{\em Tol 1214-277} - We excldued a marginally-resolved source at
approximately right ascension 12:17:17.7 and declination -28:02:56
from the 24 and 70~$\mu$m measurements, as this is likely to be a
background galaxy.  However, the source became blended with Tol
1214-277 at 160~$\mu$m, so we do not report 160~$\mu$m flux density
measurements for Tol 1214-277.

{\em II Zw 40} - The 160~$\mu$m image contains only a few square
arcmin of background.  The 160~$\mu$m background appears to contain a
signficant surface brightness gradient, which may be expected given
that the galaxy lies at a galactic latitude of $\sim -11$.
Additionally, we had difficulty reproducing the 160~$\mu$m flux
density published by \citet{eetal08}.  Given this, we did not feel
confident reporting a 160~$\mu$m flux density for this source.

~

{\em \noindent Notes on the HRS data}

{\em NGC 4356} - The galaxy falls near a 24~$\mu$m artefact we
describe as also affecting the NGC~4472 data (see below).  However,
the feature appears relatively faint and broad in the viscinity of
NGC~4356, and so we treat it as part of the background.

{\em NGC 4472} - The 24~$\mu$m image in the scan map data from AORs
22484480, 22484736, 22484992, and 22455248 were affected by two
streak-like regions that run roughly perpendicular to the scan map
direction.  These features do not appear in overlapping maps taken on
other dates during the mission.  We were unable to identify the origin
of this line.  All we can say is that the positions of these streaks
vary with respect to the scan leg position and that the width of the
features is variable.  One of these streak-like regions runs across
the optical disc of NGC~4472, and we masked it out before making
24~$\mu$m flux density measurements.

{\em NGC 4486} - The 160~$\mu$m data within the optical disc of
NGC~4486 were notably affected by residual striping in the images.
Two strips approximately 3~arcmin in width to the north and south of
the nucleus were affected and were masked out when the 160~$\mu$m flux
density was measured.

{\em NGC 4526} - Both the 70 and 160~$\mu$m images cover only the
central 3 arcmin of the galaxy, and the 160~$\mu$m image does not
include a section on the western side of the optical disc that is 2
arcmin in width.  However, the emission is relatively centralised, so
these problems may not significantly affect the photometry.

{\em NGC 4552} - In the 160~$\mu$m data, a cirrus feature oriented
roughly east-west can be seen crossing through the optical disc of
this galaxy.  We otherwise detect no 160~$\mu$m emission; we found no
160~$\mu$m counterparts to the 24 and 70~$\mu$m central source in this
galaxy.  Hence, we are reporting the integrated flux density as an
upper limit even though we get a $>5\sigma$ detection for the
integrated flux densty within the optical disc and we detect surface
brightness features at $>5\sigma$ level.

{\em NGC 4567/4568} - The 70~$\mu$m data near this galaxy are heavily
affected by latent image effects.

{\em NGC 4636} - This is an elliptical galaxy with an optical disc
with a size of $6.0 \times 4.7$ arcmin \citep{detal91}.  At 160~$\mu$m,
we detect multiple off-center point sources within the optical disc of
the galaxy that are approximately half the brightness of the central
source and that do not appear to correspond to structure within the
galaxy.  We assume that the central source is associated with the
galaxy and the off-central sources are background galaxies, but
masking out the off-central sources was equivalent to masking out the
equivalent of most of the optical disc.  We therefore perform a
160~$\mu$m measurement within a circle with a diameter of 80 arcsec
and then apply the multiplicative aperture correction of 1.745 given
by \citet{setal07} for a 30 K source (which, among the spectra used to
calculate aperture corrections, is the closest to the expected
spectrum for this object).

{\em NGC 4647/4649} - While the optical disc of these two galaxies
overlap, NGC 4649 produces relatively compact 24~$\mu$m emission and no
detectable 70 or 160~$\mu$m emission.  We assume that the optical disc
of NGC~4647 contains negligible emission from NGC~4649. Hence, we are
able to report separate flux densities for each source at 24~$\mu$m,
flux densities for NGC~4647 at 70 and 160~$\mu$m, and upper limits for
the 70 and 160~$\mu$m flux densities for NGC~4649 using the part of
NGC 4649 that does not include NGC~4647.  Also, the 70~$\mu$m image is
strongly affected by latent image artefacts.

{\em NGC 4666} - This galaxy was observed in photometry map mode.  The
galaxy is observed in such a way that the latent image removal in the
24~$\mu$m data processing leaves a couple of NaN values near the
center of the galaxy.  These pixels correspond to locations between
peaked emission, so it is clear that the data are not related to
saturation of the detectors.  We interpolated over these pixels before
performing photometric measurements.

\subsection{Comparisons of photometry to previously-published results}

The MIPS calibration at this point is very well established, and
comparisons between MIPS and IRAS photometry have already been
performed \citep{eetal07}.  Therefore, we believe that the most appropriate
check of our photometry would be to compare our measurements to other
published MIPS photometry measurements.  As indicated above, MIPS
photometric measurements have previously been published for a
significant fraction of the data that we used.  While it is
impractical to cite every paper that has been published based on the
MIPS data for these galaxies, three papers have published MIPS data
for significant subsets of galaxies in the SAG2 and HeViCS samples.
We use these papers to check our data processing.

\subsubsection{Comparisons with SINGS data}
\label{s_photometry_singscomp}

SINGS was a survey with all of the {\it Spitzer} instruments that
observed a cross-section of a representative sample of galaxies within
30~Mpc.  A total of 15 galaxies from the SAG2 surveys and in HeViCS
were originally observed with MIPS in SINGS.  Preliminary photometry
for the survey was published by \citet{detal05}, while the final
photometry was published by \citet{dggetal07}.  We compared our data
to the data from \citet{dggetal07}.  However, we exclude NGC~5194/5195
because we are reporting one set of measurements for the system while
Dale et al. report separate flux densities for each galaxy.

The ratio of the \citet{dggetal07} 24~$\mu$m flux densities to ours is
$0.97 \pm 0.08$, which is very good.  The largest outlier is NGC 6822,
where we measure a $\sim30$\% higher flux density than Dale et al.
However, as we indicated above, this is a galaxy that is large in
angular size and that has infrared emission that extends outside its
optical disk.  Additionally, the emission from foreground cirrus
structure is relatively strong compared to the diffuse emission from
the galaxy itself.  Ultimately, this may be a case where measuring the
diffuse emission from the target galaxy is simply frought with
uncertainty.  Aside from this case, however, the comparison has
produced very pleasing results.

In comparing the \citet{dggetal07} 70~$\mu$m flux densities to our
own, we found one galaxy with a factor of $\sim5$ difference in the
flux densities.  This was NGC~4552, an elliptical galaxy with
relatively weak emission from a central source.  Dale et al. reported
a flux density of $0.52 \pm 0.11$ Jy for this galaxy, which is a
factor of 5 higher than our measurement.  The Dale et al. number could
be a factor of 10 too high because of a typographical error; when we
measured the flux density the SINGS Data Release 5 (DR5)
data\footnote{Available at \\
http://data.spitzer.caltech.edu/popular/sings/20070410\_enhanced\_v1/
.} using the same apertures that we used for our data, we obtained
$0.04 \pm 0.02$~Jy.  This measurement from the SINGS data is a factor
of 2 lower than the measurement from our mosaic.  However, our image
of this galaxy was made using both SINGS data and additional 70~$\mu$m
data that was taken after the SINGS photometry was published, and so
the measurement from our new mosaic may be more reliable.

At 160~$\mu$m for NGC~4552, we reported an upper limit that is a
factor of $\sim1.5$ lower than the \citet{dggetal07} measurement. Again, we
think our measurement could be more reliable because we combined SINGS
data with other scan map data not available to Dale et al., and so the
signal-to-noise in our data should be better.

Excluding NGC~4552, the ratio of the \citet{dggetal07} 70~$\mu$m flux
densities to ours is $1.11 \pm 0.07$.  At 160~$\mu$m, the ratio of the
Dale et al. flux densities to ours is $1.20 \pm 0.07$.  This shows
that some systematic effects cause the Dale et al. measurements to be
slightly higher than ours, although the agreement is close to the
calibration uncertainty of the data, and the scatter in the ratios is
very small.  

If Dale et al. used the data in DR5, then their 160~$\mu$m
measurements would have been based on data in which the flux
calibration factor is 5\% higher than the one we used, which could
explain part of the discrepancy at 160~$\mu$m.  However, this does not
completely explain the discrepancy, and since the flux calibration
factor in the SINGS DR5 70~$\mu$m data is the same as ours,
differences in the factor cannot explain the discrepancies in that
wave band.  Although we used data not available to Dale et al. to
produce some of our images, we still see the systematic effects in the
cases where we used exactly the same data as SINGS, so differences in
the data used should not lead to differences in the photometry.

One possible cause for the systematic offsets in the photometry could
be the differences in the way the short term drift was removed.  The
other possible cause is differences in the way flux densities were
measured and handled.  While we used relatively large apertures (1.5
times the D$_{25}$ isophote) to measure flux densities, Dale et
al. used the D$_{25}$ isophotes as apertures and then applied aperture
corrections.  To check whether the data processing was the primarily
culprit for the discrepancy, we downloaded the SINGS DR5 data and
performed photometry on those data using the same software and
apertures that we had applied to our own (after correcting the
160~$\mu$m flux calibration to match ours).  The ratio of the
measurements from the SINGS DR5 data to the measurements from our data
is $0.95 \pm 0.07$ at 70~$\mu$m and $1.08 \pm 0.04$ in the 160~$\mu$m
data.  This shows that the measurement techniques are responsible for
a significant part of the systematic offsets between the Dale et
al. measurements and ours, while the data processing differences
probably cause an additional offset in the 160~$\mu$m data.

Overall, we are satisfied with how our measurements compares to the
data from \citet{dggetal07}.  The scatter in the measurements is
relatively small when difficult cases are excluded.  The remaining
differences are at levels that are comparable to the calibration
uncertainties and that are in part related to the measurement
techniques, and these differences probably reflect limitations in the
photometric accuracy that can be achieved with MIPS data for nearby
galaxies in general.

\subsubsection{Comparisons with \citet{eetal08} data}

\citet{eetal08} published data a survey of starburst galaxies that
spanned a broad range of metallicities.  22 of the 66 galaxies overlap
with the SAG2 sample: 21 of the galaxies are in the DGS, and NGC~5236
is in the VNGS.  Although Engelbracht et al. applied colour
corrections while we have not, it is still useful to compare the data.

The ratio of the \citet{eetal08} 24~$\mu$m measurements to our
24~$\mu$m measurements is $1.00 \pm 0.13$, indicating that our
measurements agree with the Engelbracht et al. to within 13\%.
However, this includes some infrared-faint galaxies where both
Engelbracht et al. and we report $>10$\% uncertainties in the flux
density measurements.  If we use data where the 24~$\mu$m flux
densities from both datasets are $>0.1$~Jy, the ratio becomes $1.00
\pm 0.05$.  The remaining dispersion is equivalent to the uncertainty
in the flux calibration, which is very good.

Engelbracht et al report 24~$\mu$m flux densities for two objects for
which we do not report flux densities.  For Tol 0618-402, we have
reported an upper limit of 0.0015~Jy, while Engelbracht et al. have
reported a $\sim 4 \sigma$ detection ($(4.4 \pm 1.2) \times
10^{-4}$~Jy).  We are reporting $<5\sigma$ detections as upper limits,
so, given the signal-to-noise in the Engelbracht et al. measurement,
we would not report a flux density for this galaxy.  None the less,
our upper limit for Tol 0618-402 is consistent with the Engelbracht et
al. flux density.  The other object is NGC~5253, for which we reported
no flux density measurement because the 24~$\mu$m emission originates
from an unresolved source that saturates the 24~$\mu$m array.
Engelbracht et al. report a flux density for this galaxy but made no
special notes about it.  Although the saturation may not be too
difficult to deal with when measuring the flux density, we prefer to
be more conservative and report no flux density for this object.

In comparing the \citet{eetal08} 70~$\mu$m data to ours, we found one
galaxy where the flux density measurements differ by a factor of 2.
For Tol 1214-277, our 70~$\mu$m flux density measurement is $0.073 \pm
0.010$~Jy, whereas Engelbracht et al. report $0.031 \pm 0.003$~Jy.
The signal from the source is hardly $5\sigma$ above the noise in our
image of this galaxy.  We also probably used a broader measurement
aperture than Engelbracht et al.  Engelbracht et al. used apertures
that were adjusted to radii that encompassed all pixels with emission
above a set signal-to-noise level, whereas we used a 3~arcmin diameter
aperture, which was our standard aperture for point-like sources.  Our
aperture may have included additional signal not included by
Engelbracht et al.

Excluding Tol 0618-402 (where we report an upper limit and
\citet{eetal08} report a $\sim1.5$ detection) and Tol 1214-277
(discussed above), our 70~$\mu$m flux density measurements agree well
with those from Engelbracht et al.  The ratio of the \citet{eetal08}
70~$\mu$m measurements to ours is $1.04 \pm 0.17$.  For sources above
1~Jy, where the signal-to-noise is primarily limited by the calibration
uncertainty, the ratio is $1.02 \pm 0.09$, which is comparable to the
calibration uncertainty of 10\%.

A comparison of the \citet{eetal08} 160~$\mu$m measurements with ours
(for galaxies we detected above the $5\sigma$ level and where we did
not encounter problems with photometry) does not show agreement that
is as good as for the 24 and 70~$\mu$m data.  Aside from
non-detections, the ratio of the Engelbracht 160~$\mu$m flux densities
to ours is $0.88 \pm 0.28$.  Measurements for UGC~4483 and UM~461 are
particularly discrepant.  We measure 160~$\mu$m flux densities that
are greater than a factor of 2 higher than the Engelbracht et
al. measurements.  Howver, these are very faint galaxies; the flux
densities are $<0.2$~Jy. The Engelbracht et al. measurements are at
the $<3\sigma$ level, and we used 160~$\mu$m data that would have been
unavailable when the Engelbracht et al. results were published, so the
improved signal-to-noise in our data could have allowed us to make
more accurate measurements for these faint galaxies. Excluding UGC
4483 and UM 461, the ratio of Engelbracht et al. 160~$\mu$m
measurements to ours is $0.96 \pm 0.19$.  The scatter in the ratio is
still larger than the calibration uncertainty of 12\%, but this may
reflect issues with simply measuring 160~$\mu$m flux densities in the
MIPS data for these dwarf galaxies, many of which are fainter than
1~Jy or in small fields.  Additionally, the colour correction applied
by Engelbracht et al. could have increased the dispersion in the
ratios.

Overall, this comparison has shown excellent agreement between the 24
and 70~$\mu$m flux densities measured by us and by \citet{eetal08}.
In the 160~$\mu$m data, we found two discrepancies that cause some
concern, but we think these are unique cases.  Our 160~$\mu$m flux
densities for other DGS sources were in general agreement with the
Engelbracht et al. measurements, thus demonstrating the reliability of
our data reduction and photometry for these data.

\subsubsection{Comparisons with \citet{amsetal11} data}

\citet{amsetal11} published a multiwavelength survey of 369 nearby
star-forming galaxies that includes 24~$\mu$m data.  23 of the
galaxies in the HRS and 2 of the additional HeViCS galaxies overlap
with the galaxies in the Ashby et al. sample.  Ashby et al. used
SExtractor to measure flux densities and then applied appropriate
aperture corrections, which is notably different from the aperture
photometry that we applied.

We have one galaxy where our 24~$\mu$m measurements differ notably
from \citet{amsetal11}.  For NGC 3430, we measured $0.4101 \pm 0.0164$
Jy, but Ashby et al. measured $0.17 \pm 0.01$ Jy.  We used the same
data as Ashby et al. to produce our image, so differences in the raw
data cannot explain the difference in flux densities.  An examination
of the image does not reveal any indication of any problems with
producing the image or making the photometric measurement.  The IRAS
25~$\mu$m flux density measurements of $0.27 \pm 0.04$ Jy given by the
Faint Source Catalog \citet{metal90} and $0.78 \pm 0.05$ Jy given by
\citet{ssm04} are also higher than the Ashby et al. measurements
numbers but still disagree with ours and with each other.  We
ultimately suspect that the mismatching flux densities could be
indicatinve of a problem with the Ashby et al. measurement obtained
using SExtractor for this specific galaxy, as the Ashby et
al. measurement is lower than all other measurements at this
wavelength.  Unfortunately, we do not have access to the final Ashby
et al. images and cannot make any assessment of the differences
between their image and ours, which would help us to understand the
problem further.

Excluding NGC 3430, the ratio of the \citet{amsetal11} measurements to
ours is $0.90 \pm 0.09$.  Ashby et al. assume that their uncertainties
are 8\%, so the dispersion in the ratio of measurements is reasonably
good.  The systematic offset may be a consequence of differences
between the flux density measurement methods.  The second largest
mismatch between our measurements and the measurements from Ashby et
al. is for NGC~4688, a late-type galaxy with significant diffuse, low
surface brightness 24~$\mu$m emission; Ashby et al. measure a flux
density $\sim30$\% lower than ours for this galaxy.  Ashby et al. also
noted differences between the flux densities measured for NGC~4395 by
themselves and by \citet{detal09}, which they thought could be the
result of  incorrectly measuring diffuse emission in NGC~4395 using
SExtractor.  We suspect that this could also be the reason for the
mismatch between the flux density measurements for NGC~4688 and may be
the reason for the $\sim10$\% offset in flux density measurements
between the reported flux densities from their catalog and ours.

\section{Summary}

We have gathered together raw MIPS 24, 70, and 160~$\mu$m MIPS data
for galaxies within the SAG2 and HeViCS surveys and reprocessed the
data to produce maps for the analysis of these galaxies.  We have also
performed aperture photometry upon the galaxies in the surveys that
can be used to study the global spectral energy distributions of these
sources.  The flux density measurements and the images will be
distributed to the community through the Herschel Database in
Marseille\footnote[18]{Located at http://hedam.oamp.fr/ .} so that the
broader astronomical community can benefit from these data.

As tests of our data processing and photometry, we have performed
comparisons between our photometric measurements and measurements
published by \citet{dggetal07}, \citet{eetal08}, and
\citet{amsetal11}.  Our measurements generally agree well with the
measurements from these other catalogs, and we have documented and
attempted to explain any major discrepancies or systematic offsets
between their measurements and ours.  Given the good correspondence
between our measurements and the measurements from these other
surveys, we are confident about the reliability of our photometry
measurements.

\section*{Acknowledgments}

We thank Laure Ciesla, Ali Dariush, Aur\'elie Remy, and Matthew
W. L. Smith for their assistance with either identifying data for
galaxies within the {\it Spitzer} archive, evaluating the final images
and photometry, or proofreading the manuscript.  We also thank the
anonymous reviewer for his/her comments.  GJB is funded by the STFC.
This research has made use of the NASA/IPAC Extragalactic Database
(NED) which is operated by the Jet Propulsion Laboratory, California
Institute of Technology, under contract with the National Aeronautics
and Space Administration.

{}

\label{lastpage}

\end{document}